\documentclass[%
  aip,
rsi,%
 amsmath,amssymb,
preprint,%
]{revtex4-1}
\usepackage{amsmath}
\usepackage{amssymb}
\usepackage{graphicx}
\usepackage{dcolumn}
\usepackage{bm}
\usepackage{listings}

\usepackage{xcolor}

\begin{document}
\title{Search and Return Model for Stochastic Path Integrators}
\author{J. Noetel}
\affiliation{Department of Physics, Humboldt-University at Berlin,
  Newtonstr. 15, D-12489 Berlin, Germany}
\affiliation{National Institute for Space Research 12227-010 Sao Jose dos Campos, Brazil}

\author{V. L. S. Freitas}
\affiliation{Federal University of Sao Paulo 12247-014 Sao Jose dos
  Campos, Brazil}

\author{E. E. N. Macau}
\affiliation{National Institute for Space Research 12227-010 Sao Jose
  dos Campos, Brazil}
\affiliation{Federal University of Sao Paulo 12247-014 Sao Jose dos
  Campos, Brazil}

\author{L. Schimansky-Geier}
\affiliation{Department of Physics, Humboldt-University at Berlin,
  Newtonstr. 15, D-12489 Berlin, Germany}
\affiliation{Department of Physics and Astronomy, Ohio University,
  Athens, OH 45701, USA}
\affiliation{Berlin Bernstein Center for Computational Neuroscience,
  Humboldt University at Berlin, Unter den Linden 6, D-10099 Berlin, Germany}  %

\begin{abstract}
We extend a recently introduced prototypical stochastic model
describing uniformly the search and return of objects looking for new
food sources around a given home. The model describes the kinematic
motion of the object with constant speed in two dimensions. The angular
dynamics is driven by noise and describes a ``pursuit'' and ``escape''
behavior of the heading and the position vectors. Pursuit behavior
ensures the return to the home and the escaping between the two
vectors realizes exploration of space in the vicinity of the given
home. Noise is originated by environmental influences and during
decision making of the object. We take symmetric $\alpha$-stable noise since
such noise is observed in experiments. We now investigate for the
simplest possible case, the consequences of limited knowledge of the
position angle of the home. We find that both noise type and noise
strength can significantly increase the probability of returning to the
home. First, we review shortly main findings of the model presented in
the former manuscript. These are the stationary distance distribution
of the noise driven conservative dynamics and the observation of an
optimal noise for finding new food sources. Afterwards, we generalize
the model by adding a constant shift $\gamma$ within the interaction
rule between the two vectors. The latter might be created by a permanent
uncertainty of the correct home position. Non vanishing shifts
transform the kinematics of the searcher to a dissipative
dynamics. For the latter we discuss the novel deterministic properties
and calculate the stationary spatial distribution around the home.
\end{abstract}

\date{\today}%
\maketitle
\textbf{
There exists search at global scales and local search centered around a 
given home position. In the latter case, the searcher does not only look 
for a new target but is also required to permanently return to the home 
position. Such behavior is typical for many insects and achieves 
technical importance for self-navigating robotic systems. We propose a 
stochastic nonlinear model for local search which does not distinguish 
between the two aims. The dynamics bases on an unique pursuit and escape 
behavior of the heading from the position vector realizing thereby 
optimal exploration of space and the return to the home. We discuss the 
mechanics of the searcher and inspect the role of noise. Such randomness 
is present in the decision making rule of selecting the new heading 
direction. We consider Levy noises with different degree of 
discontinuity and report about steady spatial densities for the 
searchers. Also we report about an optimal noise intensity that a 
searcher finds a target at nearby places where the spatial distribution 
of the searcher is maximal.
}
\section{Introduction}

The usage of different kinds of noise in global search problems is
very popular \cite{Bennichou2011}. For non-local search applications of
L{\'e}vy Flights and of various types $\alpha$-stable white noise have
proven\cite{Klages} to be very effective. However, applications of
such kinds of noise in models of local search are very rare. Local
search reduces  the exploration of the neighborhood of a certain
localization. The use of power law distributed step length to explore the neighborhood of a certain spot seems to be counterproductive. 

Such local search is observed for objects which are able to keep track
of distances and orientations as they move and use these informations
to calculate their current position in relation to a fixed location.
The latter can be a nest or a source of food and is called home
\cite{Mittelstaedt}. If the position of the home and the angle towards
the home are known then the method is called path integration
\cite{Mittelstaedt,Cheng,Wang,Klages} and the specific exploration
behavior might be based on some internal storage mechanism
\cite{Seelig,Green} or external cues \cite{Zeil} as for example
special points of interest or pheromonic traces, etc.

In our model studied below we orientate on studies and experiments
concerned with ants, bees and
flies\cite{Wehner_et_al_1996,ElJundi_2017,Kim_Dickinson_2017}. For
these animals such homing behavior is found. Especially, the various
two dimensional spatial patterns which the local searchers draw during
their motion have received a lot of interest
\cite{Wehner_1981,Wehner_et_al_1996,Vickerstaff,Waldner_2018}. 
A similar deterministic kernel of the heading 
dynamics as considered below was introduced earlier by Vickerstaff and 
Di Paolo \cite{Vickerstaff_2005} based on 
Mittelstaedts' investigations \cite{Mittelstaedt_bico}. But
their profound mathematical explanation based on principles of
nonlinear and stochastic dynamics seems to be still a challenge.

In a broader sense, the better understanding of local search problems
is of high technical interest. Nowadays a new age of spatial
exploration \cite{chien_2017} has started, in which robots are
reaching places that human beings have never approached, with explorer
robots being projected for missions in ocean
\cite{Hook_2013,Girdhar_2011,Leonard_et_al_2007,Dubowsky_2005}, space
\cite{chien_2017,Dubowsky_2005}, etc. Such applications as, for
example, an autonomous spacecraft or search and rescue missions
require autonomy since the vehicles must be able to make decisions and
cannot be idle while waiting for directions.

Also homing behavior and local search in engineering applications are
of relevance to autonomous navigation in systems of surveillance, data
collection, exploration, monitoring, etc. Autonomous vehicles
\cite{Leonard_et_al_2007,Duarte_et_al_2016} might move inside an area,
around a chosen point, sometimes visit it and return to the local
search using internal cues instead of external ones \cite{Nirmal,
  Moeller}.


In a recent manuscript we introduced a novel stochastic model for local
search\cite{Noetel_2018}.  Here, we extend the investigations on this
stochastic model, that considers an active particle with constant
speed in two dimensions. The constant speed is common in a variety of
models \cite{mikhailov,Romanczuk} and also observed in experiments
with insects \cite{Kim_Dickinson_2017}.

Our model aims to mimic the motion of simple organisms,
self-navigating automata or other moving objects which shall explore
the surrounding space and return to its home. We implement the local
search around the home via a coupling term between the heading angle
of the particle and the angle formed by its current position and the
home. The coupling is defined in an uniform way for the search epoch as
well as for the return one. Hence we do not distinguish between the
two aims.

We assume that the searcher possesses an internal storage
mechanism. In a Cartesian frame of reference two angles are stored
while in the reference frame of polar coordinates the spatial distance
and the angle between the heading direction and the direction towards
the home needs to be known. We point out, that in our model no knowledge
of the distance towards the home is required in a Cartesian frame of reference. This is in contrast to similar 
path integrators \cite{Vickerstaff_2005}. However, in the reference frame of polar
coordinates the particle would perform path integration. The resulting
spatial motion allows the particle to explore the vicinity of the home
in order to find new food sources and the searchers consecutively
return to the home.


In section \ref{sec:model}, we briefly review the model. In sections
\ref{sec:det_model} and \ref{sec:model_noise} we report on central
findings in \cite{Noetel_2018}. The presentation of the latter in the
current manuscript is necessary for the understanding of the novel
aspects considered below. This discussion concentrates on results with
noise present.

In the next section \ref{sec:extension_gamma} we extend
our investigation and consider a situation with a limited knowledge of
the position of the home. The escape-pursuit aligning mechanism between the
two vectors is weakened by introducing a shift in the interacting
rule. As a consequence the motion of the searcher starts to become
dissipative and collapses to a limit cycle in the two dimensional
space. We calculate the approximative spatial densities for large and
small noise intensities.  Finally, we summarize our findings in
section \ref{sec:concl}.

Special attention is paid to the influence of the random decision
process of the searcher. The noise therein originates from an
uncertainty in the definition of heading direction for the active
particle. This uncertainty is related to neural processing and
decision making of the new direction. Reasons are in a vagueness about
the exact position of the home or in a limited capability to choose
an exact direction of motion due to external perturbations.

In many studies on stochastic nonlinear dynamics, Gaussian white noise
is applied as source of dynamical randomness. However, we model the
noise as symmetric $\alpha$-stable white noise source which also
includes the case of a Gaussian noise with the particular choice
$\alpha=2$. Inclusion of white L{\'e}vy noise in this local search
problem was stimulated by findings in experiments with the fruit fly
Drosophila Melanogaster \cite{Kim_Dickinson_2017}. The authors of this
study report on a L{\'e}vy-kind statistics in the angular dynamics
during the local search of the fruit fly. Despite those randomness
acting with heavy tails, the wrapping of the  dynamic angles onto the
interval $(-\pi,\pi]$ will allow asymptotically the establishment of a
continuous steady state distribution of the searchers around the
home. As will be elaborated below, the latter is centered around home
without the introduced shift. Otherwise, searchers accumulate
probability in a crater-like shape above the stochastic limit cycles
if the shift hinders precise orientation.

\section{The Search and Return Model}
\label{sec:model}

\begin{figure}[h]
  \includegraphics[width=0.39\linewidth]{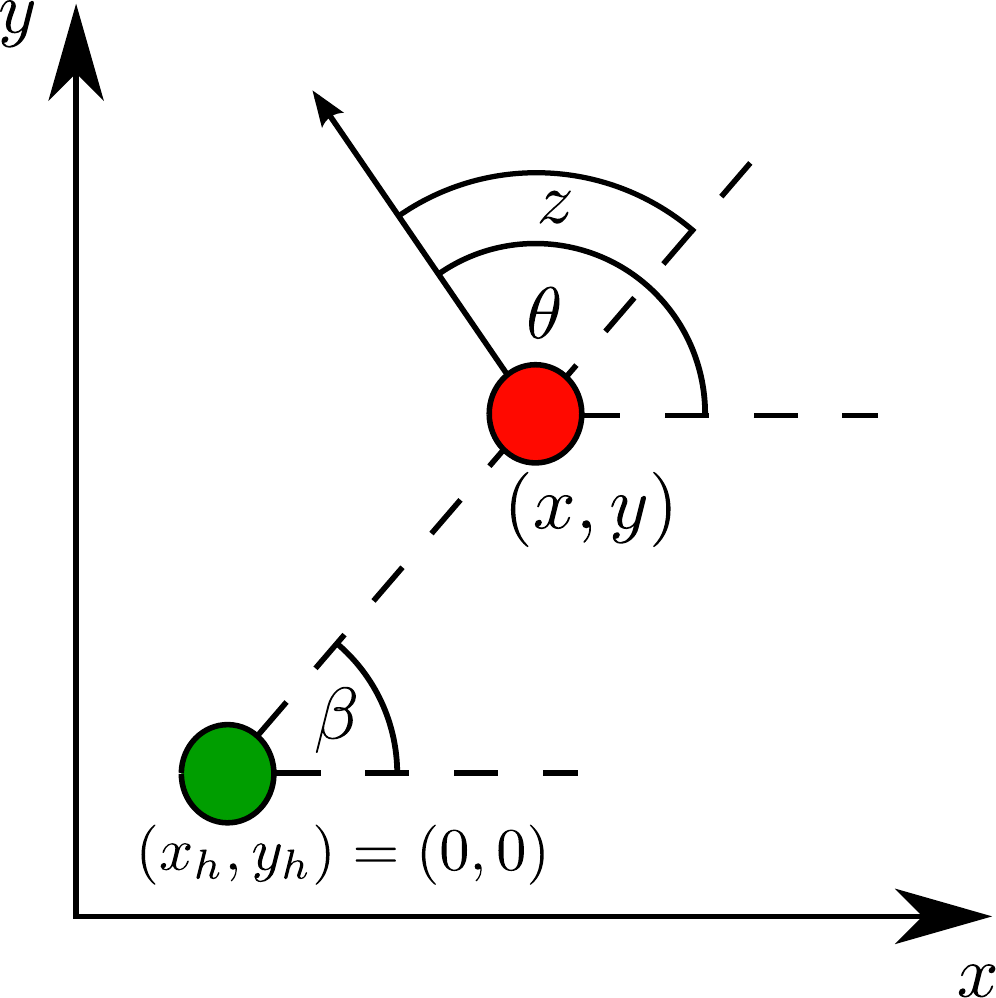}
    \caption{Schematic introduction of position and heading vector
      with angles $\beta(t)$ and $\theta(t)$. The difference between
      the two angles is assigned by $z(t)=\theta(t)-\beta(t)$. The
      searching object is colored red and moves in two dimensions with
      the Cartesian position $x(t)$ and $y(t)$.  The home possesses
      color green and is situated for convenience at the origin. The
      distance of the searcher from the home is
      $r(t)=\sqrt{x(t)^2+y(t)^2}$.}
    \label{fig:beta}
\end{figure}

We consider an active Brownian particle \cite{Romanczuk} whose
position vector is given by $\vec{r}(t)=\{x(t),y(t)\}$. This position
vector reads in polar coordinates $\vec{r}(t)=r(t)
\{\cos(\beta(t)),\sin(\beta(t))\}$ with distance $r(t)$ and direction
$\beta(t)$. We assume that the searcher moves with constant speed
$v_0$
\begin{eqnarray} 
\dot{\vec{r}}=\vec{v}(t)=v_0
\begin{pmatrix}
         \cos\theta(t) \\ \sin\theta(t)
        \end{pmatrix}\,,
\label{eq:r_dot}
\end{eqnarray}
$\dot{\vec{r}}$ denotes the temporal derivative of the position vector
$\vec{r}$ and $\theta$ is the heading angle of the particle pointing
along the velocity vector as is depicted in Figure
\ref{fig:beta}.

Since the home is situated at the origin of the Cartesian reference
frame, the position vector $\vec{r}$ points always out of the home towards the particle's 
current position. The corresponding angle
$\beta(t) \in[0,2\pi)$ is given by:
\begin{equation}
\beta(t) = \arctan{\frac{y(t)}{x(t)}}, 
\label{eq:beta}
\end{equation}
as also sketched in Figure \ref{fig:beta}. When arriving the home
$x(t_a)=y(t_a)=0$, both vectors are anti-parallel
$\beta(t_a-0)=\theta(t_a)+\pi$. Inertia would require that the object
leaves the home with $\beta(t_{a}+0)=\theta$. Inspired by the behavior of the fruit flies, we discuss at the end of the search study in 
Sect.\eqref{subsec:meanfirsthit} a random reset of the heading in close proximity to the home. 

The search and return dynamics is part of the evolution rule for the
heading angle $\theta(t) \in[0,2\pi)$. We require that $\dot{\theta}$ 
evolves in time as
\begin{equation}
\dot{\theta} =  \kappa \sin(\theta-\beta) + \frac{\sigma}{v_0} \xi(t).
\label{eq:dottheta}
\end{equation}
The first item is the deterministic drift term of this equation. It
might be motivated by a escape and pursuit behavior
\cite{romanczulsgcouszin2009} of the position and velocity
vectors. For positive values of the parameter $\kappa$, if $\theta(t)
\in \{\beta(t)-\pi/2,\beta+\pi/2\}$, the velocity vector is repelled
from the position vector. It is the situation where the velocity
vector also points out of home and the object still leaves the
home. Thereby, a separation of the two vectors guarantees a larger
exploration of the space in the vicinity of the home. Otherwise, if
$\theta(t) \in \{-\beta(t)-\pi/2,-\beta+\pi/2\}$ the velocity vector
is pointing oppositely to the position vector. In this situation the searcher
is already on the way home. To find the position of the home an
alignment, i.e. an attraction of the two vectors, is preferable as
described by Eq.(\ref{eq:dottheta}).

In the second term on the r.h.s. of Equation \eqref{eq:dottheta}
$\xi(t)$ stands for a source of symmetric $\alpha$-stable white noise.
The simulations were performed considering an Euler integration step for the deterministic drift
and the noise is implemented following \cite{Weron_95,Weron_book,Nolan}. For details see Appendix \ref{app:numeric}. 
It serves as an uncertainty in the definition of the heading direction
at time $t$. It might be caused by a limited knowledge of either the
heading direction, itself, or of the angle between the current
position of the particle and the home $\beta$. Generally, it stands
for the decision making step of the searcher to choose a new
direction. The noise strength is $\sigma$. In the case of $\alpha=2$,
increments of the angle are uncorrelated in time and with Gaussian
support.White noise with $\alpha < 2$ in the angular dynamics yields a
continuous description for a run and tumble like motion with fast
tumbling epoch \cite{noetel:2017} as it was also found and reported in
the experimental study \cite{Kim_Dickinson_2017}.

Further on, we will cross to a dimensionless description. Let us
assign the dimensionless time as $t^\prime= \kappa t$ and the
dimensional position from the origin as $\vec{r^\prime}(t^\prime) =
\vec{r}(t^\prime)/r_c$ with $r_c= v_0/\kappa$. Afterwards one gets for
the new temporal variables
$x^\prime(t^\prime),y^\prime(t^\prime),\theta^\prime(t^\prime)$ the
dimensionless deterministic dynamics. For simplicity, we reassign
$t^\prime \rightarrow t, x^\prime(t^\prime) \rightarrow x(t), y^\prime(t^\prime) \rightarrow
y(t), \theta^\prime(t^\prime)\rightarrow \theta(t)$ and get
\begin{eqnarray}
  &&\frac{\rm d}{{\rm d}t}\,x = \cos(\theta)\,,\\ 
 &&\frac{\rm d}{{\rm d}t}\,y = \sin(\theta)\,,\\
  &&\frac{\rm d}{{\rm d}t}\,\theta = \sin(\theta-\beta) + \frac{\sigma^{\prime}}{v_0} \xi(t)\,,
\end{eqnarray} 
with $\beta(t)= \arctan(y(t)/x(t))$. Notable, the deterministic dynamics
depends only on the dimensionless initial coordinates. We get the dimensionless equations from the outgoing ones by selecting simply $\kappa=1$ and $v_0=1$.  
In contrast, the noise intensity has to be rescaled and reads $\sigma^{\prime} =\sigma/\kappa^{1/\alpha}$ and the $v_0$ remains in front of the noise source in the dimensionless frame. But likewise for the other values we omit the prime and replace $ \sigma^\prime \rightarrow \sigma$. Later on in simulations, we take for simplicity values of $\kappa$ and $v_0$ always equal to unity. Another choice would imply a multiplicative rescaling of the noise intensity.

We underline that the proposed model is an uniform evolution law for
both, search and return. Our model does not artificially distinguish
between a search period and a return period as it is assumed for several observed food searchers such as ants \cite{Vickerstaff}. 
There is no switcher between the two aims and
both episodes of the evolution are described by a single function. It
is also worth to mention that other functions similarly to the
considered sine in (\ref{eq:dottheta}) are able to exhibit similar
effects as discussed, later on.


\subsection{Deterministic Dynamics}
\label{sec:det_model}
Let us shortly remind deterministic properties of the considered
model. With this purpose we put $\sigma=0$ in
(\ref{eq:dottheta}). After introducing the difference of the two
angles $z(t)=\theta(t)-\beta(t)$ and crossing to polar coordinates
with the distance $r(t)$ and the angle $\beta$ the deterministic
dynamics reads
\begin{eqnarray}
  \label{eq:dim_less}
 &&\frac{\rm d}{{\rm d}t}\,r = \cos(z)\,,\\
 &&\frac{\rm d}{{\rm d}t}\,z =  \left(1-\frac{1}{r}\right) \sin(z)\,,\\
 &&\frac{\rm d}{{\rm d}t}\,\beta = \frac{1}{r}\sin(z)\,.
\end{eqnarray} 
One immediately notices that the $\beta(t)$ evolution  is determined by the dynamics in the $r,z$-plane. Finding $r(t),z(t)$ gives the $\beta$-evolution by integration of the last equation. Therefore, the deterministic case 
reduces to the motion in the $r,z$-plane. This yields except for $z=\pm \pi$ and $z=0$ bounded oscillatory solutions which remind of trajectories within the celestrial mechanics, 
but in the considered case with constant speed \cite{Noetel_2018}. 

For $z=\pm \pi,0$ the value of $z$ (and $\beta$) does not change and the distance grows unlimited or vanishes, respectively. Approaching with $z= \pm \pi$ the value $r=0$, the trajectory jumps quickly to $z=0$ from which it grows unbounded $r(t) \to \infty$. In the stochastic case, these trajectories are without any relevance since the angular noise will change permanently the $z$-value and, hence, a constant value of $z$ will not hold. As consequence, these unbounded trajectories have zero measure in the stochastic frame.
\begin{figure}[h]
\includegraphics[width=0.41\linewidth]{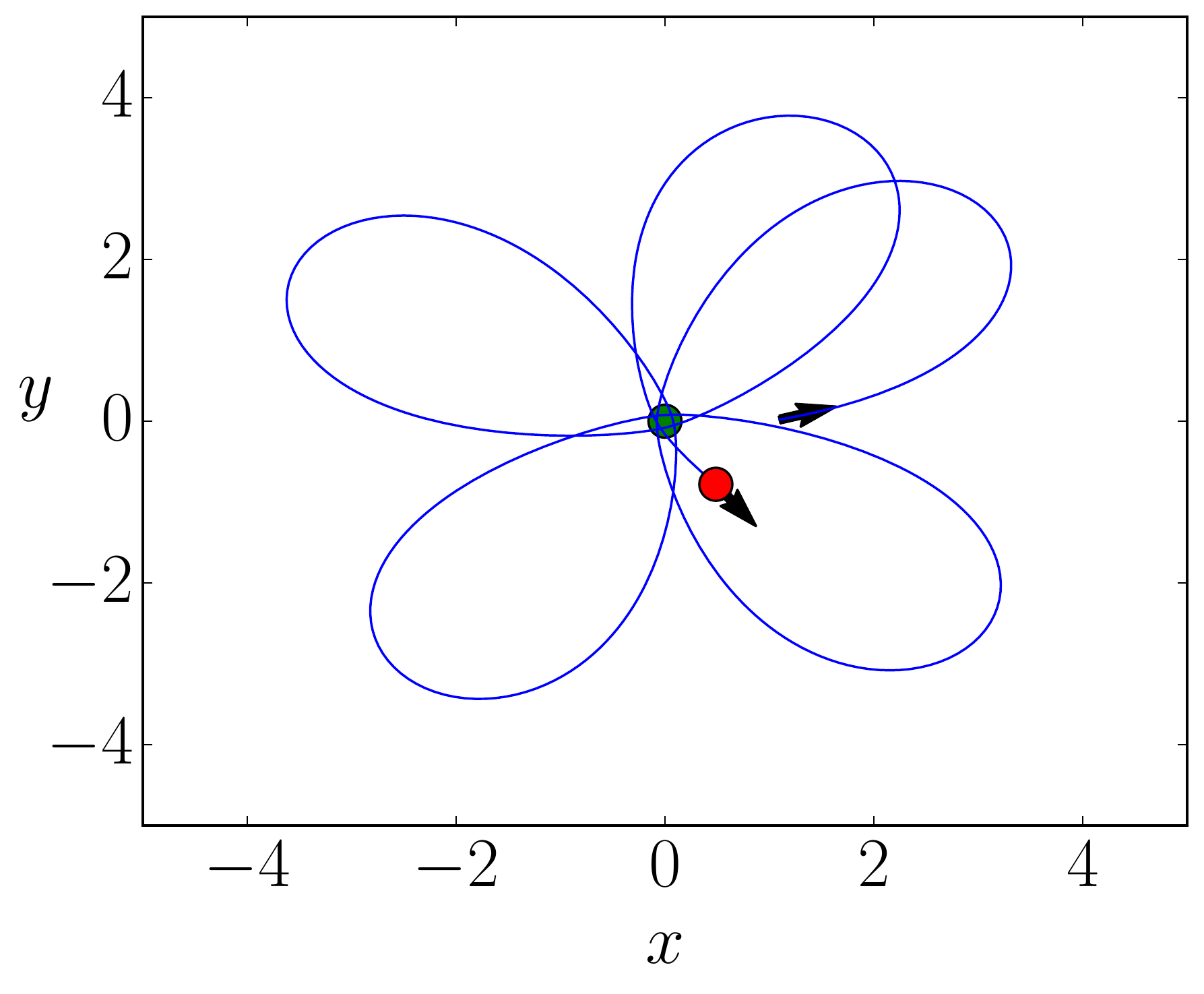}  \includegraphics[width=0.51\linewidth]{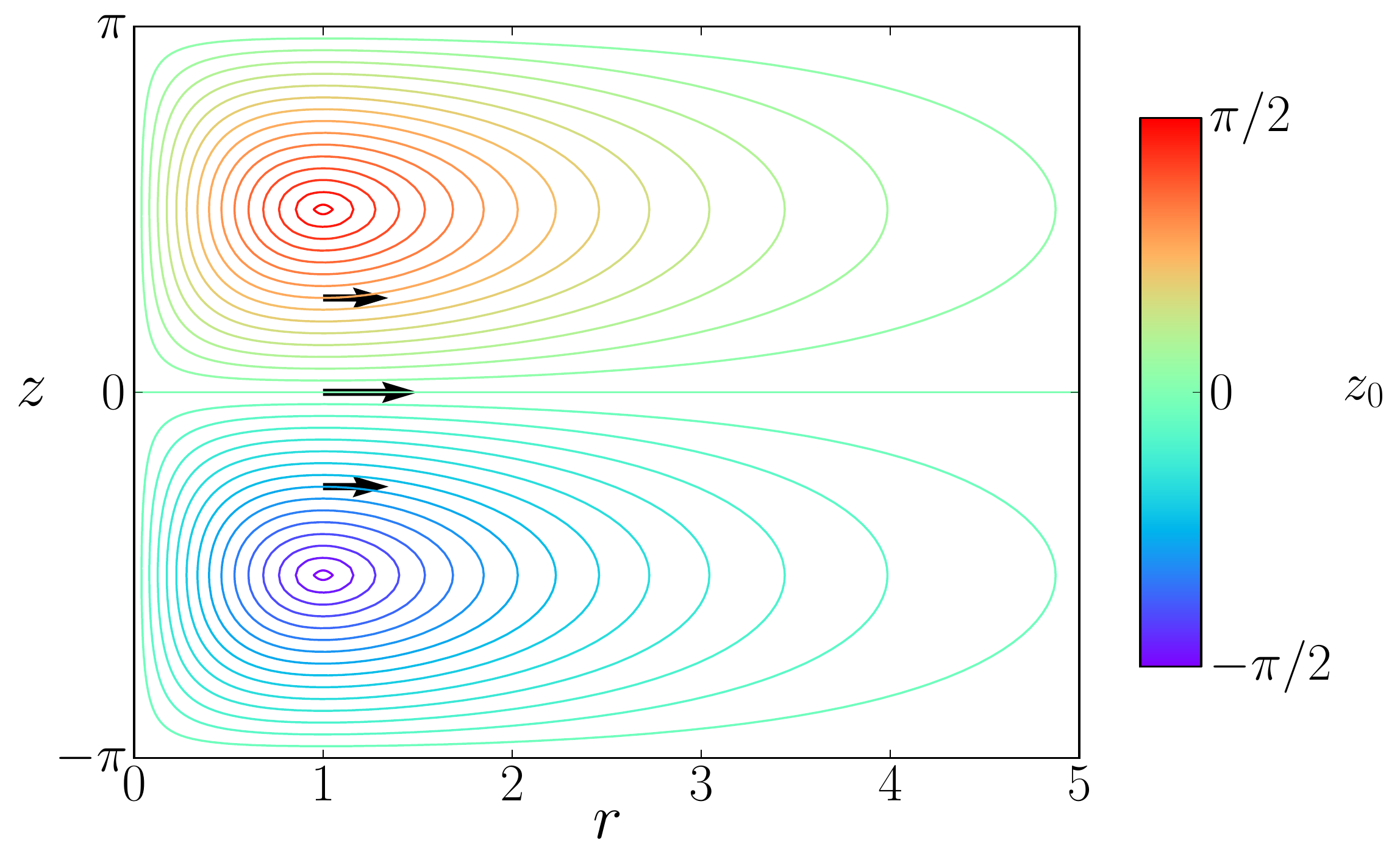}
 \caption{Left: Periodic deterministic trajectory in $x,y$ frame with anti-clockwise perihel precession during one rotation with $x_0=1,y_0=0,\theta_0=\pi/16$. Right: Trajectories in $(r,z)$ plane according to Equation \eqref{eq:sinzvr} for initial conditions $r_0=1,z0 \in ,[-\pi/2,\pi/2]$ starting at the arrow-positions. Different colors stand for different values of $X$ resulting from different initial conditions $r_0,z_0$. The green dot represents the home position. }
    \label{fig:r_z_space}
\end{figure}

All other deterministic motion is bounded. The dynamics in $r,z$ forms closed orbit around two centers at $r_0=1,z=\pm \pi/2$ with imaginary eigenvalues $\pm {\rm i}$. Every closed orbit 
possesses an integral of motion $X(r,z)=const$. To derive this the time is  eliminated considering a parametric dependence of the two variables $z(r)$. In the equation for ${\rm d }r/{\rm d}z$ the variable can be separated and we obtain by 
quadrature \cite{Noetel_2018}
\begin{equation}
X=\sin(z(r)) \exp\left(-r\right) r= \sin(z_0) \exp\left(- r_0\right) r_0 ={\rm const}.                 
\label{eq:sinzvr}
\end{equation}
The counterclockwise rotations and the anticlockwise ones, which result in dependence on the initial angle $z_0$, are presented in Fig.(\ref{fig:r_z_space}). Minimal and maximal distances of $r$ and $z$ can be easily derived. From these we can find the period of the motion $T(r_0,z_0)$ by integration of the first equation of (\ref{eq:dim_less}) with inserted integral of motion $X$ instead of $z(t)$. It takes values closely to the eigenvalue near the centers and diverges for the larger orbits. Similarly one can integrate the $\beta$-dynamics and find a precession of the perihel in the $x,y$ space during one oscillation as shown in Fig.(\ref{fig:r_z_space}). 
Also the sign of the angular momentum $L= r^2 \dot{\beta}=r \sin(z)$ does not change along one revolution since the $\sin$-function does not change 
the sign\cite{Noetel_2018}.


\subsection{The Stochastic Case}
\label{sec:model_noise} In case with noise we also cross to polar coordinates. The dynamics for the stochastic distance $r(t)$ and angle $\beta(t)$  does not change and is given by the Eqs.(\ref{eq:dim_less}). In the dynamics for the stochastic angle difference $z(t)$ following Eq.(\ref{eq:dottheta}) a stochastic source term with $\xi(t)$ appears. It becomes  
  \begin{equation}
  \label{eq:z_stoch}
\dot{z}= \left(1 -\frac{1}{r} \right)\sin(z) + \frac{\sigma}{v_0}\,\xi(t)\,.
\end{equation}   
The noise acts on the angular dynamics, only. Particles are still moving with constant speed. The noise induces a population of all possible values of $z$ and smears the motion along all $X$ values in the $r,z$-plane. The unbounded motion becomes unlikely in the stochastic system and the searcher remains with high probability at finite distances from the home. Hence, the noise stabilizes the motion.
\begin{figure}[h]
    \includegraphics[width=0.4\linewidth]{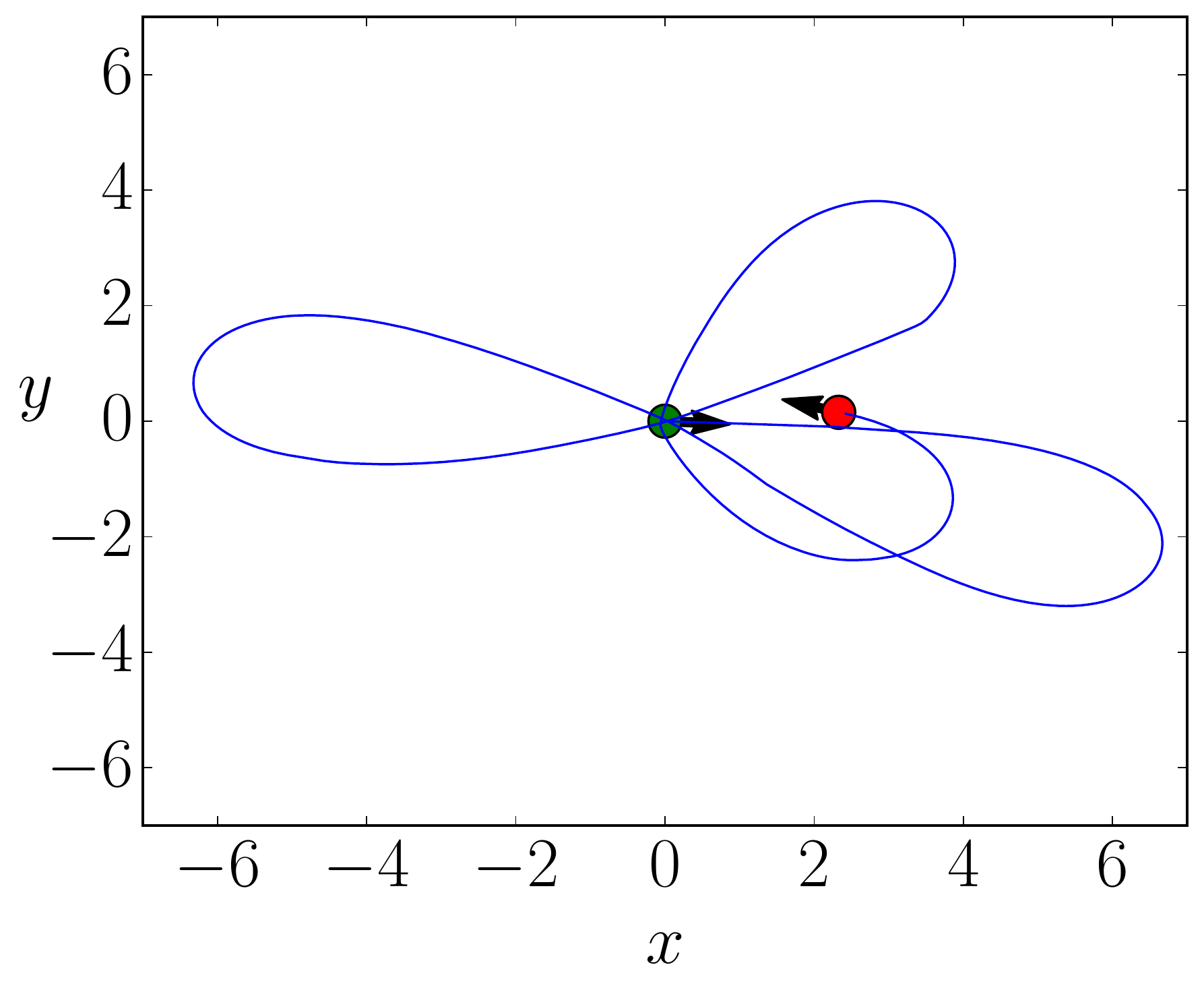}
    \includegraphics[width=0.4\linewidth]{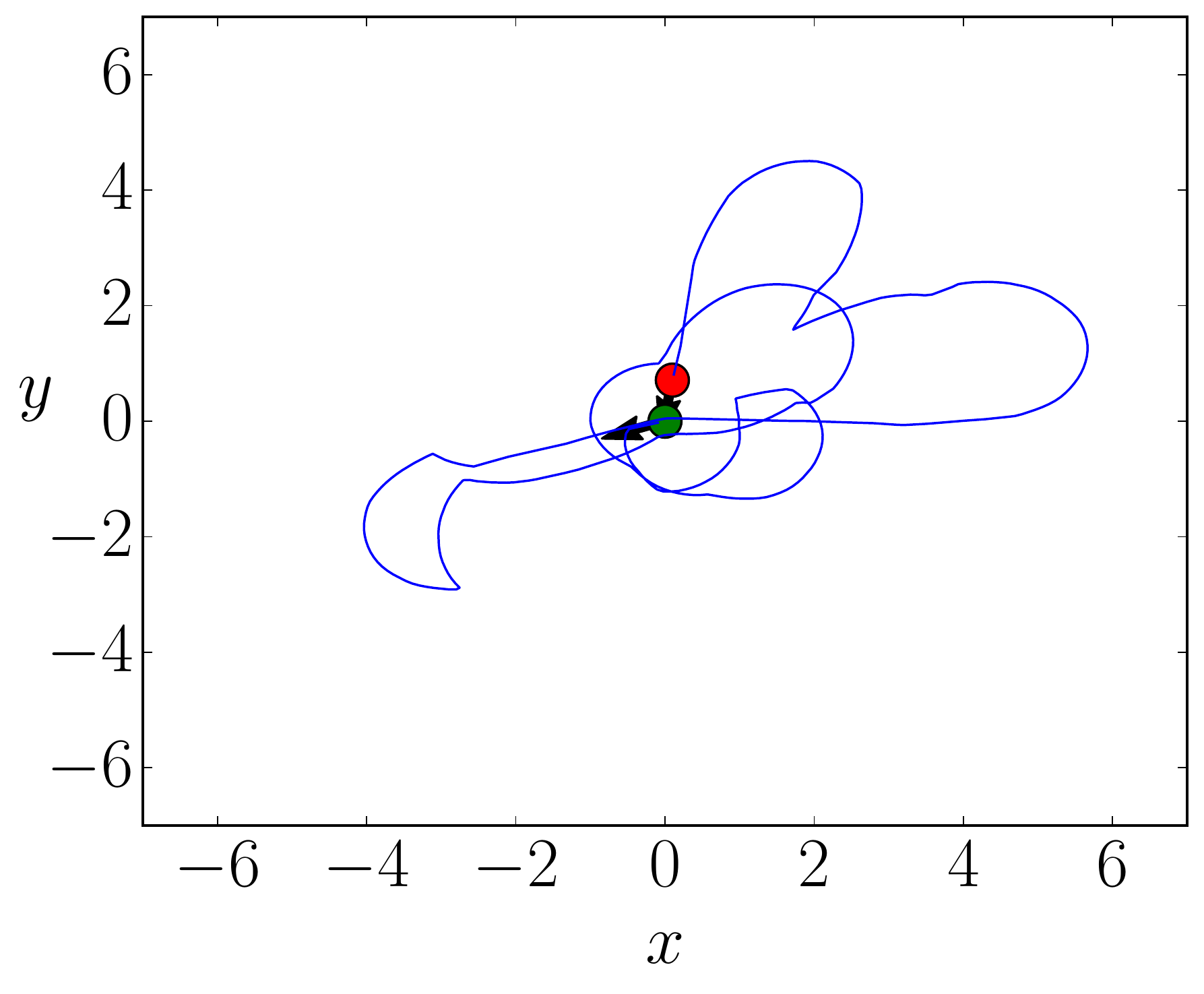}\\
      \includegraphics[width=0.4\linewidth]{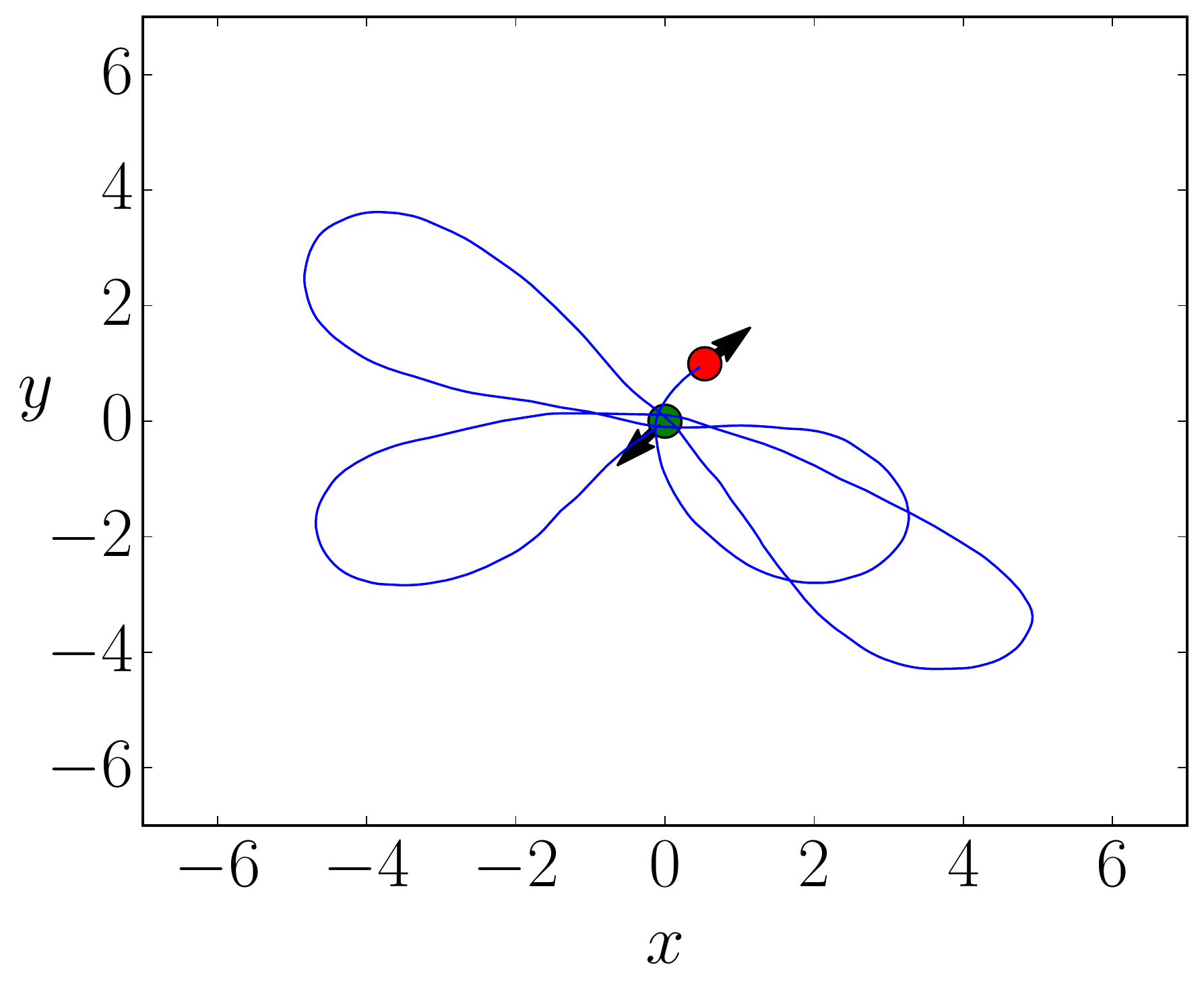}
    \includegraphics[width=0.4\linewidth]{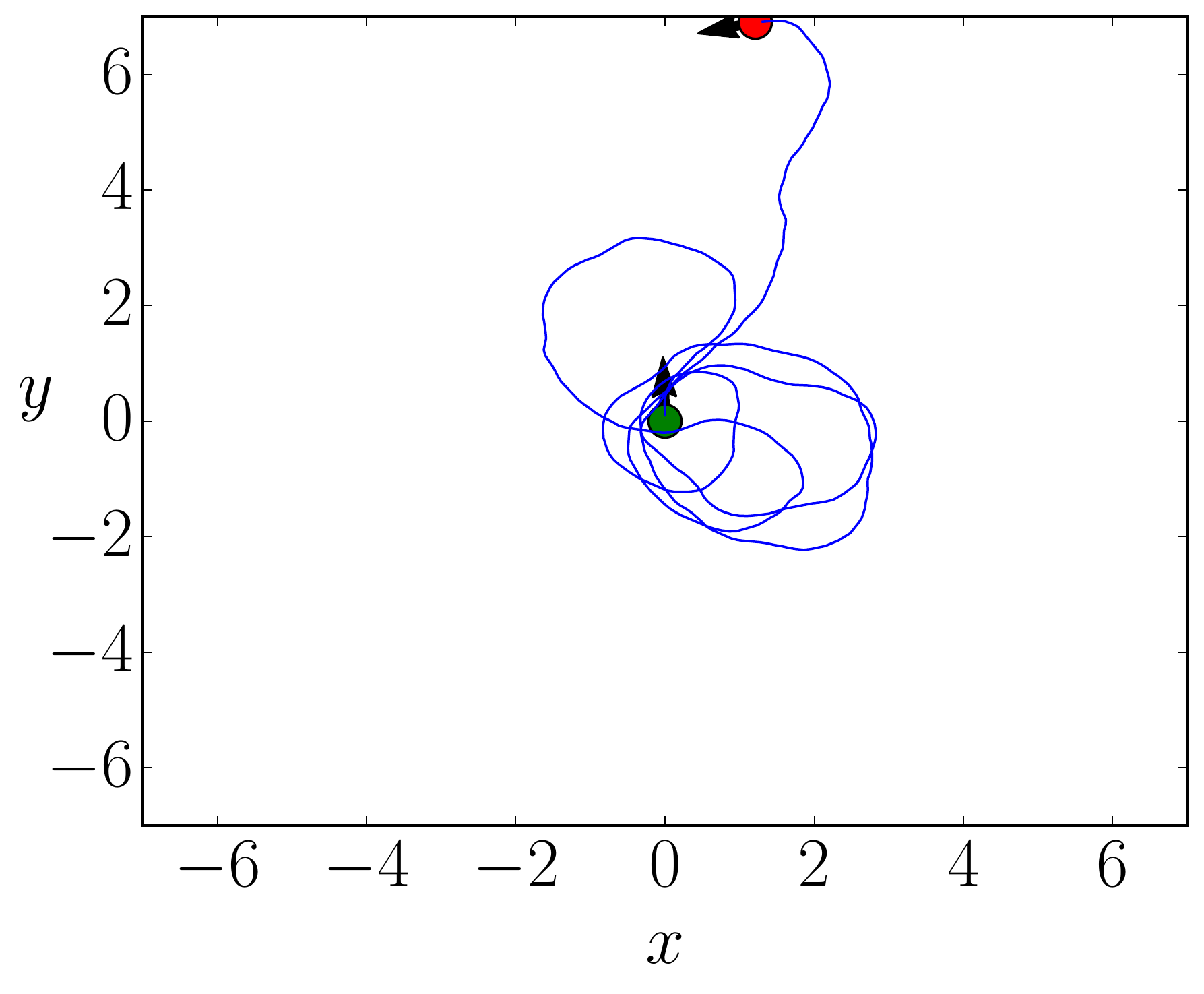}
    \caption{Sample stochastic trajectories with different noises in $(x,y)$ plane. Particle starts at $(0,0)$. 
    Upper row: $\alpha=1$; Lower row: $\alpha=2$.  Left: $\sigma^\alpha=0.01$; Right:$\sigma^\alpha=0.1$. Other parameter $v_0=1$.   }
    \label{fig:x_y_plane_highnoise}
\end{figure}

In the stochastic case one has now two more parameters. One is
$\alpha$ that characterizes the noise type. In addition we have $\sigma$,
standing for the noise strength. We will report that the noise
strength plays an important role for finding a new food source. Less important in our case is the
noise type. The noise acts on the angle which is a $2\pi$-periodic
variable. Hence the long tails of $\alpha$-stable noise are always
wrapped on the bounded interval of possible $z$-values and do not
influence significantly the stochastic evolution after relaxation to
an uniform distribution in $z$.

Typical stochastic trajectories can be seen in Figure \ref{fig:x_y_plane_highnoise}. 
The particles started at the home $(x,y)=(0,0)$ and moved for the time interval $\Delta t= 50$. 
In the left column for small noise $\sigma^\alpha=0.01$ the deterministic drift mostly determines the motion of the particle, while 
on the right for larger noise $\sigma^\alpha=0.1$ the underlying deterministic part of the motion can practically no longer be recognized. 
The noise causes a diffusion in space of possible deterministic trajectories. As derived below in the next subsection (\ref{sec:spat_dist}) 
the stationary distance density is independent of the noise type $\alpha$ and the noise strength $\sigma$.


\subsubsection{Stationary Density of Distances}
\label{sec:spat_dist}
An important measure for the behavior of the stochastic searcher is the stationary spatial distribution of the searchers. For this purpose, we introduce  transition probability
density function $P(r, z, t|r_0, z_0, t_0)$ in dependence of the distance
$r$ and angle difference $z$. This  density obeys the corresponding Fokker-Planck equation (FPE) which, 
according to \cite{Ditlevsen, Schertzer, Schertzer_Chechkin} reads in the considered case
\begin{equation}
\label{eq:FP}
  \frac{\partial}{\partial t}P = -\frac{\partial}{\partial r}\cos(z)P+\frac{\partial}{\partial z}\left(\frac{1}{r}-1\right)\sin(z)P +\left(\frac{\sigma}{v_0}\right)^\alpha \frac{\partial^\alpha}{\partial |z|^\alpha}P\,,
\end{equation}
wherein
\begin{equation}
\frac{\partial^\alpha}{\partial |z|^\alpha}P(z)=-\frac{1}{2\pi}\int_{-\infty}^\infty\,{\rm{d}k}|k|^\alpha\,\exp\left(-ikz\right)\,P(k)
\end{equation}
stands for the $\alpha$th symmetric Riesz-Weyl derivative, with $P(k)$ being the Fourier transform of $P(z)$ in the variable $z$. 
The drift term is due to the deterministic dynamics for the distance and the angle difference whereas the last term describes the influence of the $\alpha$-stable noise source. We remind that $r$ is non-negative and $z$ is $2\pi$-periodic.

\begin{figure}[h]
    \includegraphics[width=0.45\linewidth]{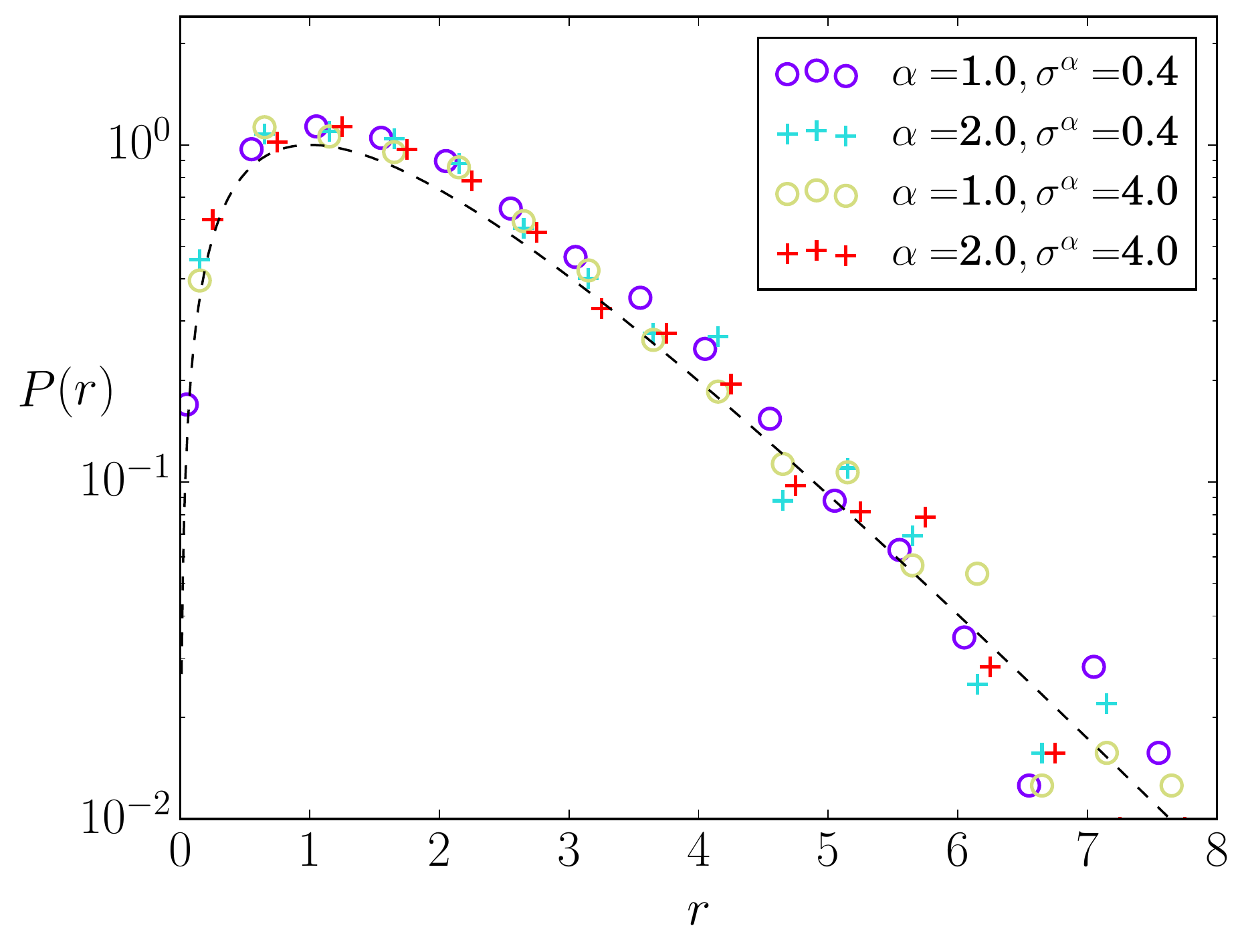}
        \caption{The stationary marginal radial pdf  $P_0(r)$ from Eq.(\ref{eq:p0}) (black dashed line). Symbols from simulations of the stochastic dynamics with different types and strength of the noise.}
    \label{fig:p0}
\end{figure}

In the asymptotic stationary limit $t \rightarrow \infty$ the initial conditions are forgotten and the density becomes stationary, i.e. $P(r,z,t\to \infty|r_0,z_0,t_0)=P_0(r,z)$. Hence one might put $\partial P_0(r,z,t)/\partial t=0$. Further on , we separate 
\begin{equation}
P_0(r,z)=P_0(r|z)P_0(z).
\end{equation}
The noise distributes the probability homogeneously in $z$ since no angular direction is distinguished. 
Therefore, $P_0(z)=1/2 \pi$. 
The fractional derivative for the wrapped constant angular pdf vanishes due to the symmetry of the noise.
There is  no effective force repelling the noisy $z$-shifts and also 
the asymptotic spatial distribution does not depend on $z$. For the latter radial pdf the equation remains to be solved:
\begin{equation}
0=\left[-\frac{\partial}{\partial r}+\frac{1}{r}-1 \right]\cos(z)P_0(r)\,.
\end{equation}
Therein the noise dependent item is absent. Dropping the $cos$-function we find the stationary radial pdf as \cite{Noetel_2018}
\begin{equation}
P_0(r,z)= \frac{1}{2\pi} r\exp\left(-r \right).
\end{equation}
The marginal distribution of distances follows immediately
\begin{equation}
P_0(r)= r\exp\left(-r \right).
\label{eq:p0}
\end{equation}
It is shown in Fig.(\ref{fig:p0}) and compared with numeric simulations for different noise types. Remarkably, the spatial density of the searcher is a Rayleigh distribution which is independent of the noise. It is independent with respect to the noise type expressed by $\alpha$ as well as to the noise intensity $\sigma$. It contains the characteristic dimensionless length $r=1$ which is the length with maximal probability. 
The mean distance is given by $<r>=2$ and the spatial coefficient of variation $CV_r=\sqrt{0.5}$  characterizing the spatial uncertainty of the random search. The typical shape is presented in Fig.(\ref{fig:p0})  

The stationary spatial density is  also independent of $\beta$. Hence, in the Cartesian frame the pdf reads:
\begin{equation}
P_0(x,y)=\frac{1}{2\pi}\exp\left(-\sqrt{x^2+y^2} \right)
\end{equation}
This pdf is maximal at the home.

Further on, we solved for arbitrary $\alpha \in (0,2]$ and $\sigma \ne 0$ the FPE \eqref{eq:FP} ,with the Ansatz $P=\exp(\lambda t)e(r,z)$, for the time dependent eigenfunction

\begin{equation}
e_1(r,z)=\sin(z)r^2\exp\left(-2 r\right)\,, 
\label{eq:efewm1}
\end{equation}

with the eigenvalue $-1/\tau$. The fractional derivative governing the angle $z$ in \eqref{eq:FP} becomes when applied to the sine function, the sine 
function itself, with a minus sign in front of it. For details see appendix \ref{app:sin}.

As seen $\tau$ which is as defined as 
\begin{equation}
\label{eq:tau}
\tau = \left(\frac{v_0}{\sigma}\right)^\alpha
\end{equation}
obtains the meaning of a relaxation time of the angle difference in the
stochastic case.  After $t\gg\tau$ the spatial distribution becomes
independent of any initial orientation $z_0$.


\subsubsection{Optimal Noise for Search of a New Food Source}
\label{subsec:meanfirsthit}
In this subsection we report on the influence of the noise to discover a new food source \cite{Noetel_2018}. We fixed the speed value $v_0=1$ in the stochastic source term of Eq.(\ref{eq:z_stoch}). Then the noise strength and its type are the tunable parameters in our stochastic model. To quantify the stochastic search for the target we obtained numerically mean times after that  the searcher finds a food spot for the first time. 
We placed a target at $(x_t,y_t)=(1,0)$. We chose this distance $d=1$ from the home as at this distance the stationary spatial density becomes 
maximal for all noise types and strength.

With a given noise type and at each value of the noise strength we made $50000$ runs. The search started in the vicinity from home with radius $r_0$ of the home with random orientations. 
As the model is supposed to describe the search of living organisms, we assume the searcher has a sensing radius $r_{{\rm sens}}$.
We do not consider a spatial extension of the target, as this can be put into the sensing radius. We determined the random search times needed to sense the target.

We consider the sensing radius to be small against the length scale $r_c$ of the system, then we chose for the simulations a value 
$r_{{\rm sens}}=0.1$. 
During its stochastic oscillatory search, the trajectory generally missed the target and returned several times to the home, passing the latter one to start the next round.  After it found the target, we resetted the particle to the initial position and restarted for the next determination of the search time.     

\begin{figure}[h]
    \includegraphics[width=0.47\linewidth]{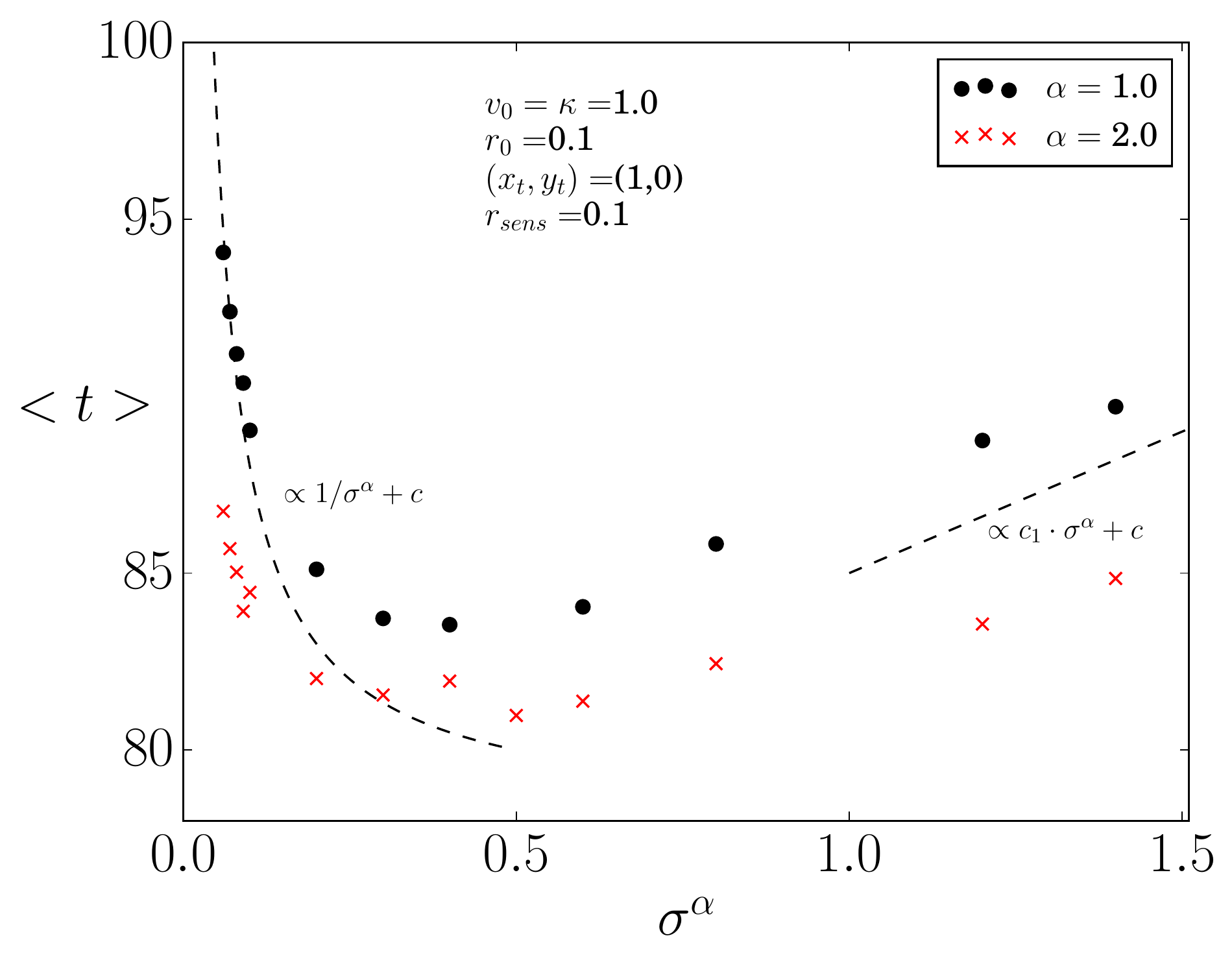}
    \includegraphics[width=0.49\linewidth]{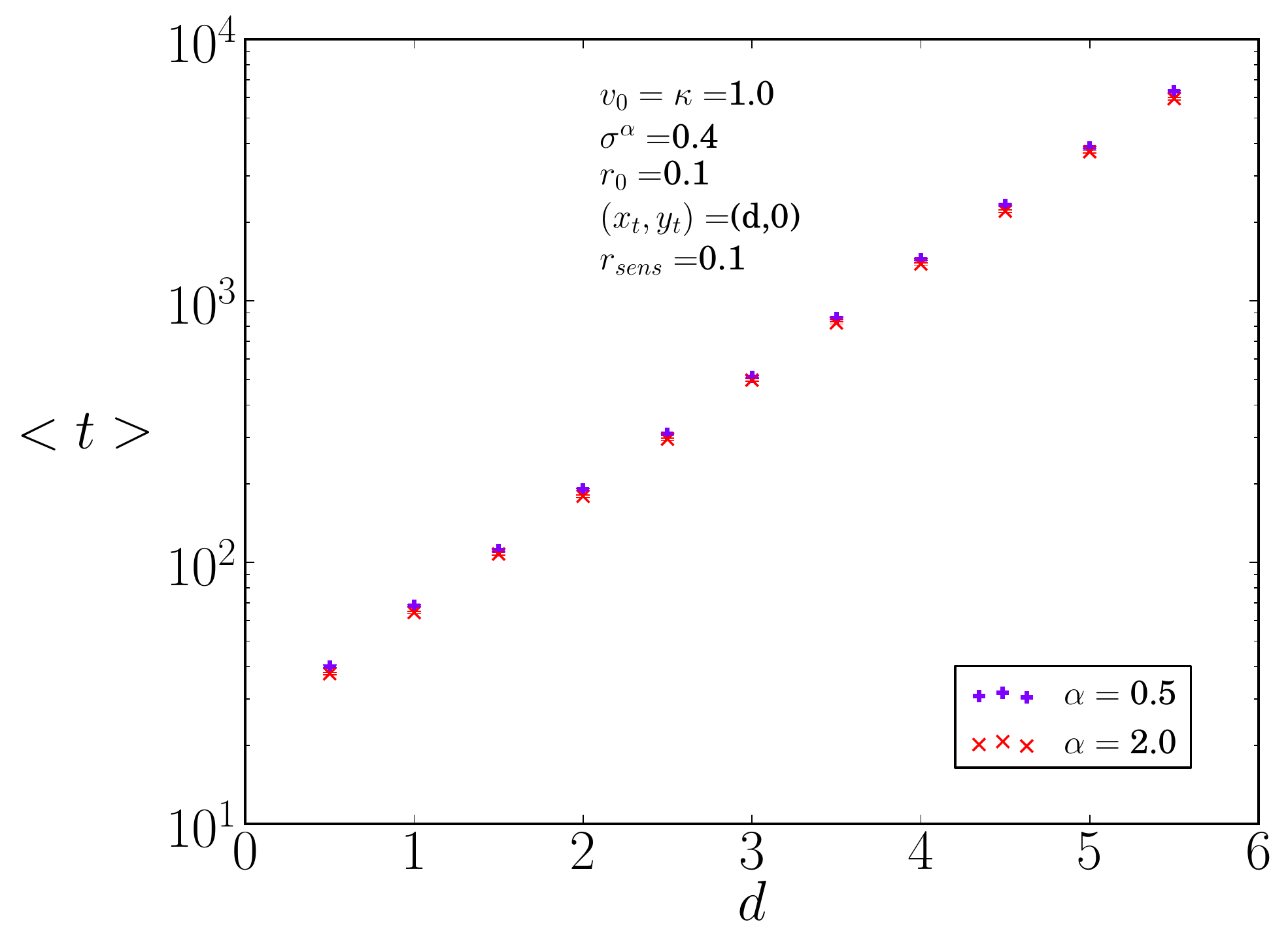}
    \caption{Left: Mean hitting time $<t>$ during search of  a new spot at a given position  $(x_t=1,y_t=0)$ found numerically in dependence of the noise intensity. Right: Mean time $<t>$ for a fixed noise intensity in dependence of the distances $d$, with $(x_t=d, y_t=0)$. }
    \label{fig:meet}
\end{figure}

Figure \ref{fig:meet} presents the mean first hitting time $<t>$ for a given food spot in dependence of the inverse relaxation time $1/\tau$. With the selected speed value,  the latter is identical to the scaled noise intensity $\sigma^\alpha$. 
First of all, the mean first hitting times are always larger than the relaxation time $\tau$ as given by Eq.( \ref{eq:tau}). However, as presented in Fig.(\ref{fig:meet}) an optimal noise strength can be seen where the mean search time becomes minimal 
\cite{Noetel_2018}. This optimal time does depend on the relaxation time $\tau$ or, respectively, on the scaled noise intensity $\sigma^\alpha$. Without noise, i.e.  $\sigma=0$, some of the deterministic trajectories will never hit, or take extremely long. For example, some existing unbounded trajectories expanding radially away from the home and those which do not hit the target during the first expansion will do it never. Hence the mean hitting time diverges in the noise-less case.  

Afterwards with low noise, the mean hitting times starts to decay nearly as $\propto C+1/\sigma^\alpha$ until reaching the minimum. Later on, for larger noise the hitting times scale with the noise intensity, i.e.$\propto \sigma^\alpha$. We also found that for larger distances to the target,  the optimal noise shifts to smaller values (not shown here). 

This non-monotonous dependence results from two counteracting effects induced by the noise. By means of the first one, trajectories will distribute over all possible orbits with different values of $X$. 
The characteristic time scale for this process is the angular relaxation time which scales as $\tau \propto 1/\sigma^\alpha$. This first effect determines the decay of $<t>$ at lower noise intensities. 

The second effect created by the noise starts to act at larger values
of the noise intensity. For these noise values, the deterministic
dynamics becomes negligible. With large noise slow diffusive search
becomes dominant. But as well known the corresponding diffusion
coefficient of free active particles scales inversely with the
noise. It decays as\cite{Milster} $D_{\rm eff}\propto 1/\sigma^\alpha.$
In consequence, the spatial relaxation needs
longer for higher noise. The time reaching in average a certain
distance $d$ in two dimensions by diffusion can be defined following
\begin{equation}
t_{\rm diff} = \frac{<d^2 >}{4 D_{\rm eff}}\propto \sigma^\alpha\,.
\label{eq:diff}
\end{equation}
On the right of Fig. \ref{fig:meet} we present the average search time for a fixed value of $\sigma^\alpha=0.4$ and different distances $d$ of the 
target. We find that the average search time grows exponentially with the distance.
\begin{figure}[h]
    \includegraphics[width=0.3\linewidth]{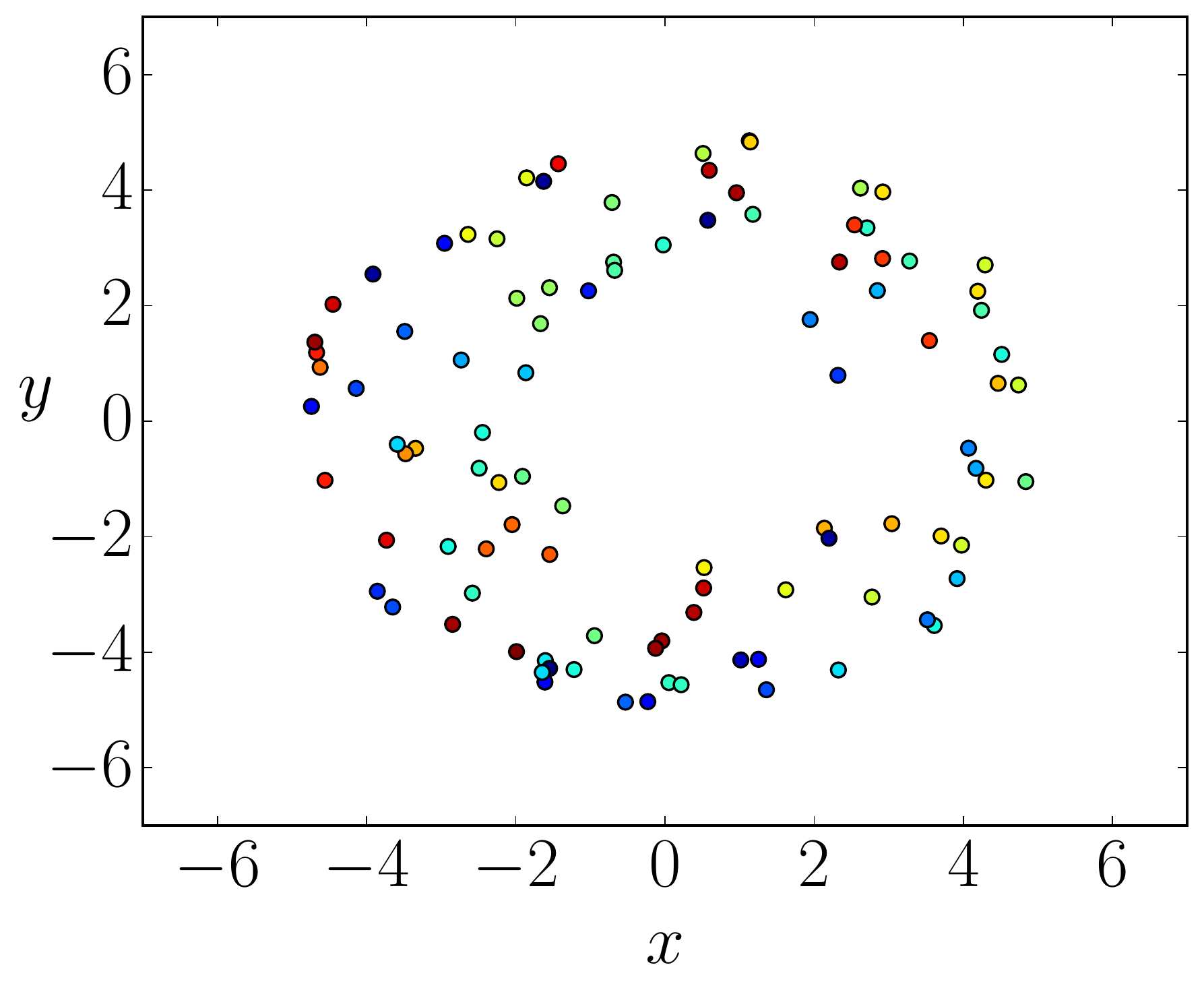}    
    \includegraphics[width=0.3\linewidth]{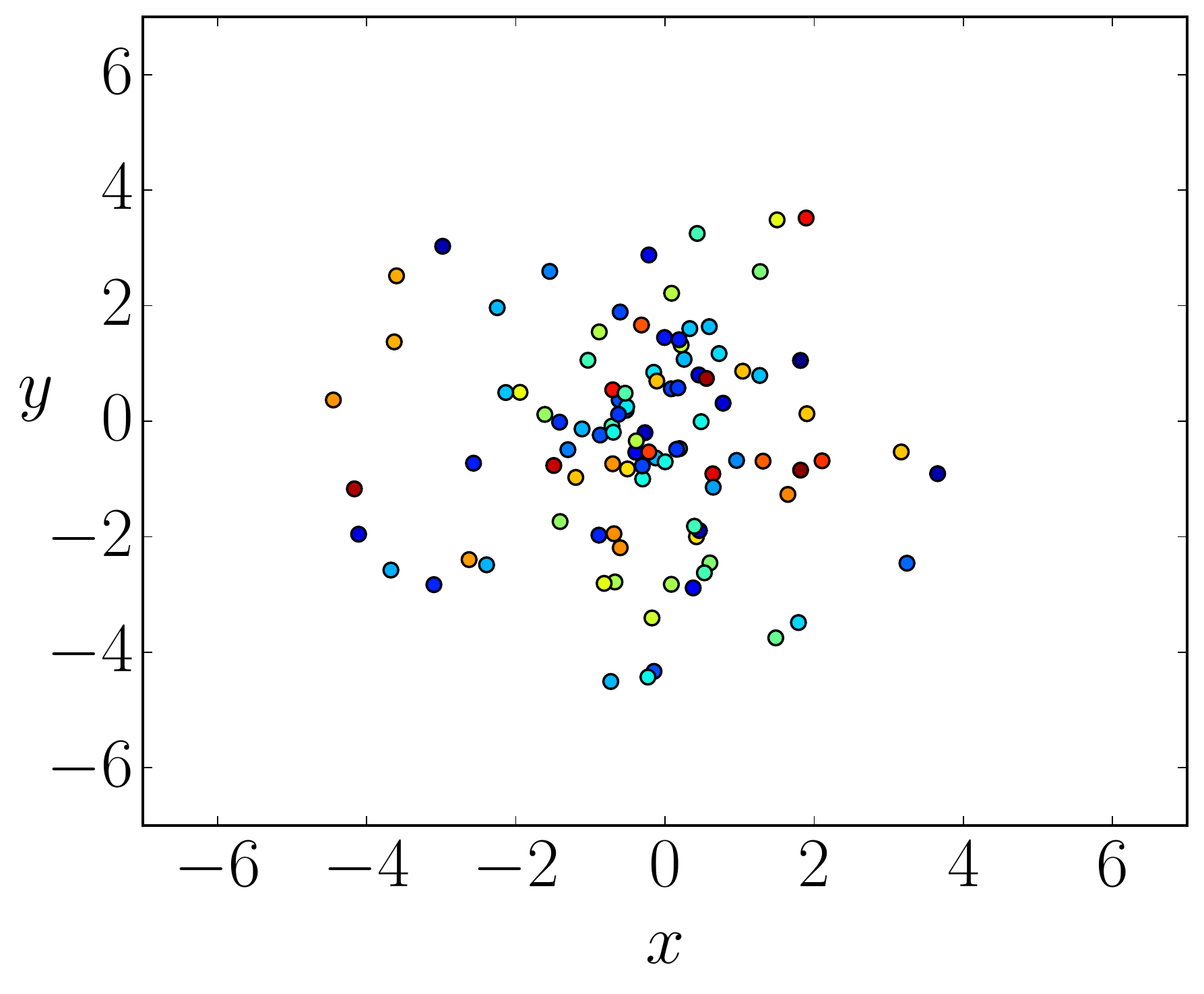}
    \includegraphics[width=0.3\linewidth]{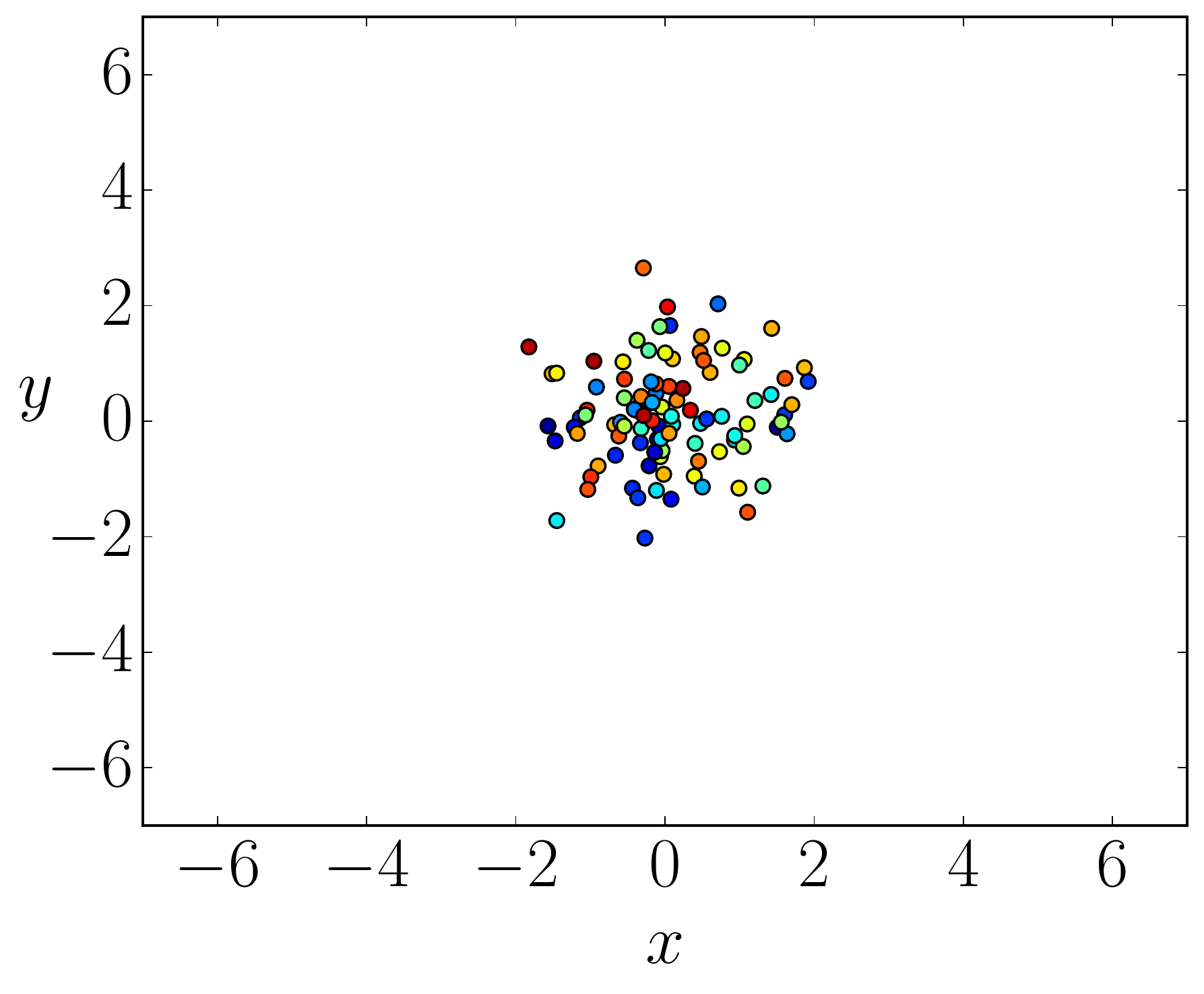}
    \caption{Spatial distribution of ensembles of $100$ searchers after simulation time $t=5$ for small $\sigma^2=0.01$, optimal $\sigma^2=0.4$ and large $\sigma^2=4$ Gaussian noise.
    Different colors of the particles correspond to different heading directions.  }
    \label{fig:ens}
\end{figure}

The different temporal
behavior of the random search is illustrated in Fig.(\ref{fig:ens})
where ensembles of $100$ stochastic searchers are presented for small,
optimal and large noise after a given time $t$. Initial conditions for
all pictures are equivalent, meaning the particles are started at a
distance $r_0=0.1$ from the home and the angles $\beta$ and $\theta$
are uniformly distributed.  In the left picture with small noise, the
angular symmetry is given due to the random initial angles. But the
single searcher follows still the deterministic dynamics along the
orbit corresponding to the particular initial state since
$t_s<\tau$. Hence, particular searchers move along the rigid
deterministic paths. Such dynamics along an orbit might take the
particle rather far away from the home. The searcher will be able in
average to find localized targets after the slow perihel precession
has slipped on most directions. In contrast, in the graph to the right
the searchers do not extend far away in space due to limited diffusive
motion. The motion of a single searcher is
undirected. Its probability has already populated uniformly in all
possible directions. The situation of the middle graph shows the desired situation. The probability density for the heading direction of a single searcher is already uniform. Searcher are fast transported in radial direction
by the underlying deterministic dynamics and noise has spread the
probability of singular searchers uniformly over all orbits. This
circumstance guarantees a successful finding of the target.

Gaussian white noise always performed slightly better in the numerical simulations. When changing $\alpha$ to lower values we observed a small increase of the mean time $<t>$.

\begin{figure}[h] 
    \includegraphics[width=0.4\linewidth]{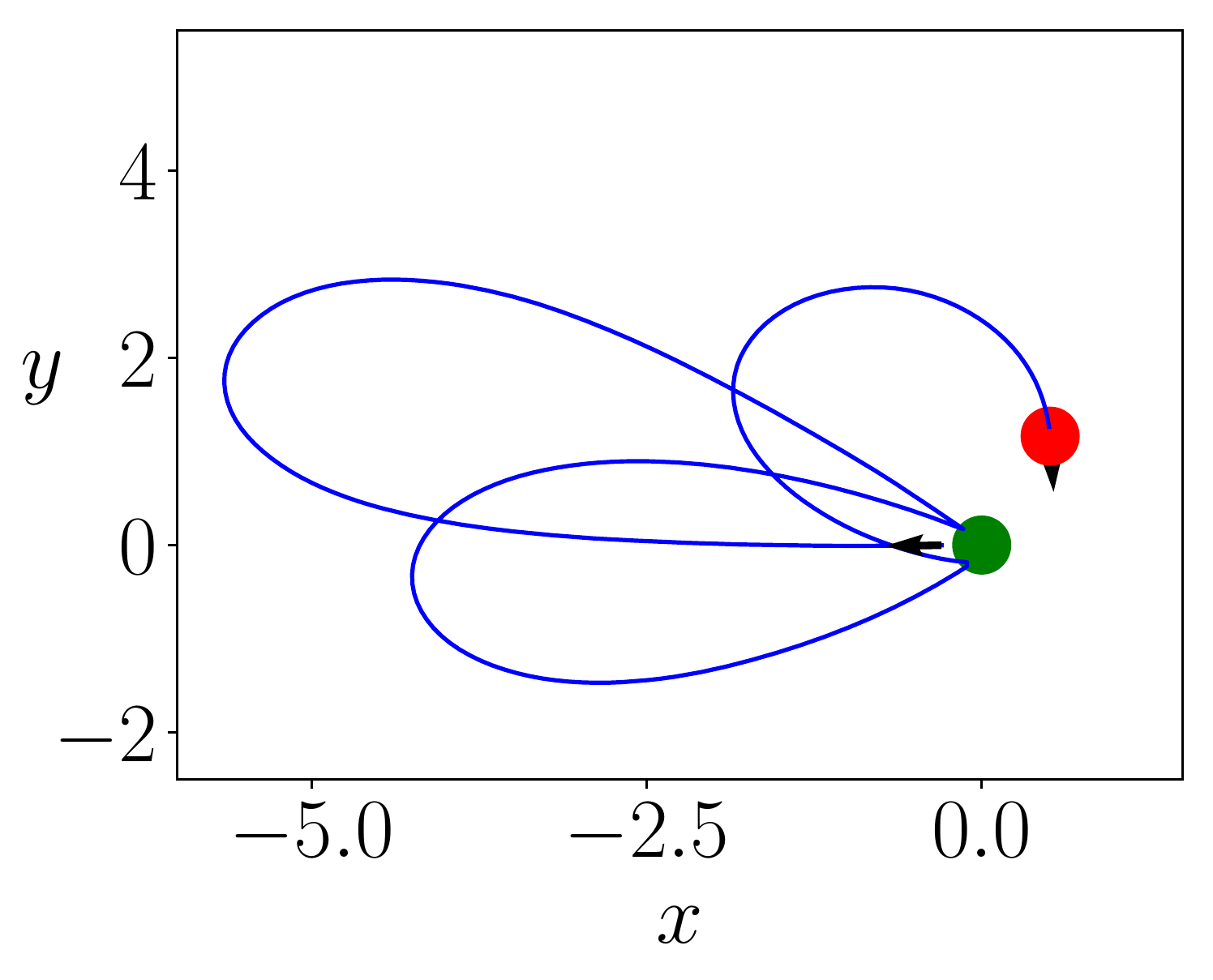}
    \includegraphics[width=0.4\linewidth]{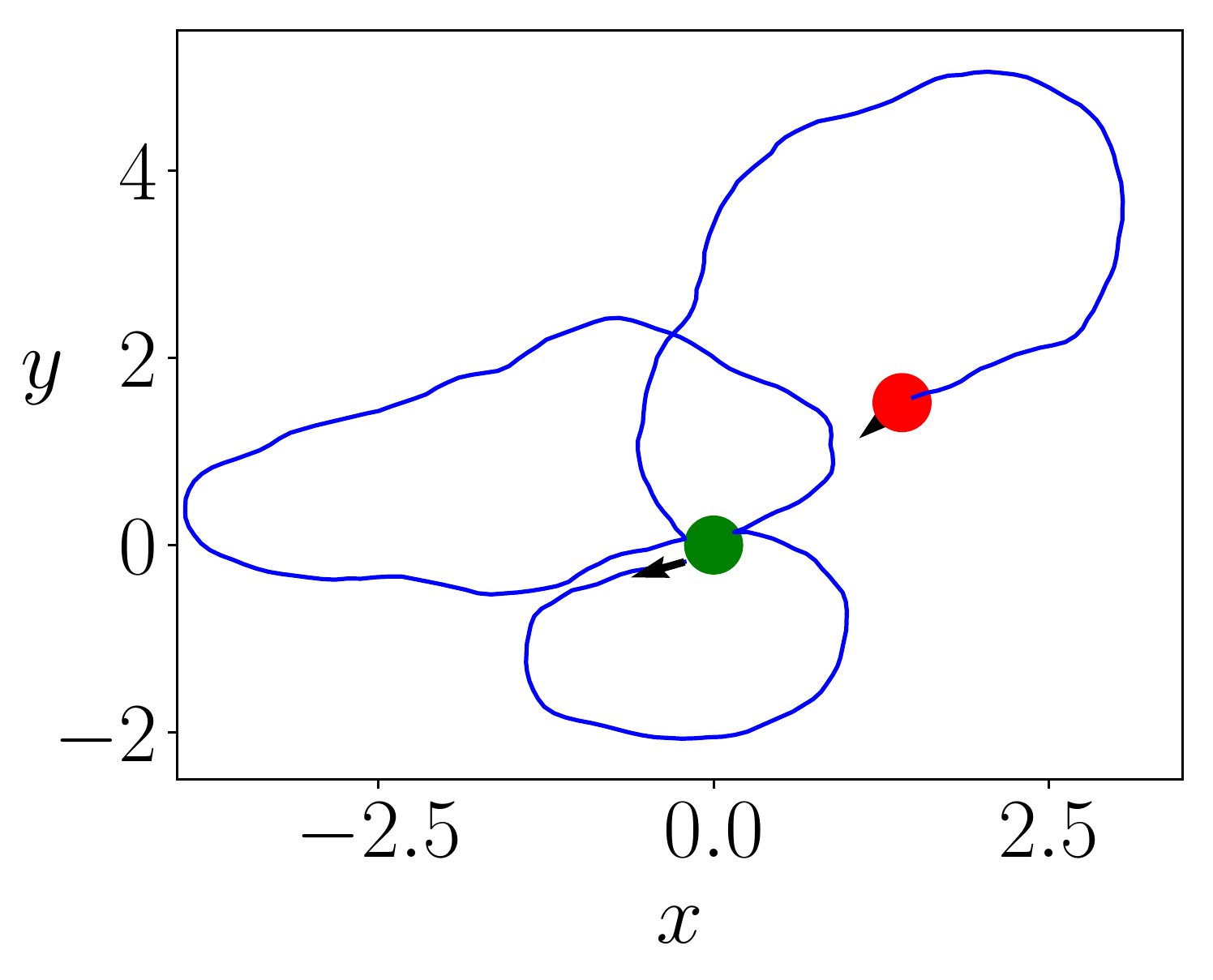}
    \caption{Sample stochastic paths with a random reset at home between two searching excursions. Left side is the  deterministic noise free case. Right graph with optimal noise $\sigma^\alpha = 0.4$. }
    \label{fig:reset}
\end{figure}

\begin{figure}[h]
    \includegraphics[width=0.49\linewidth]{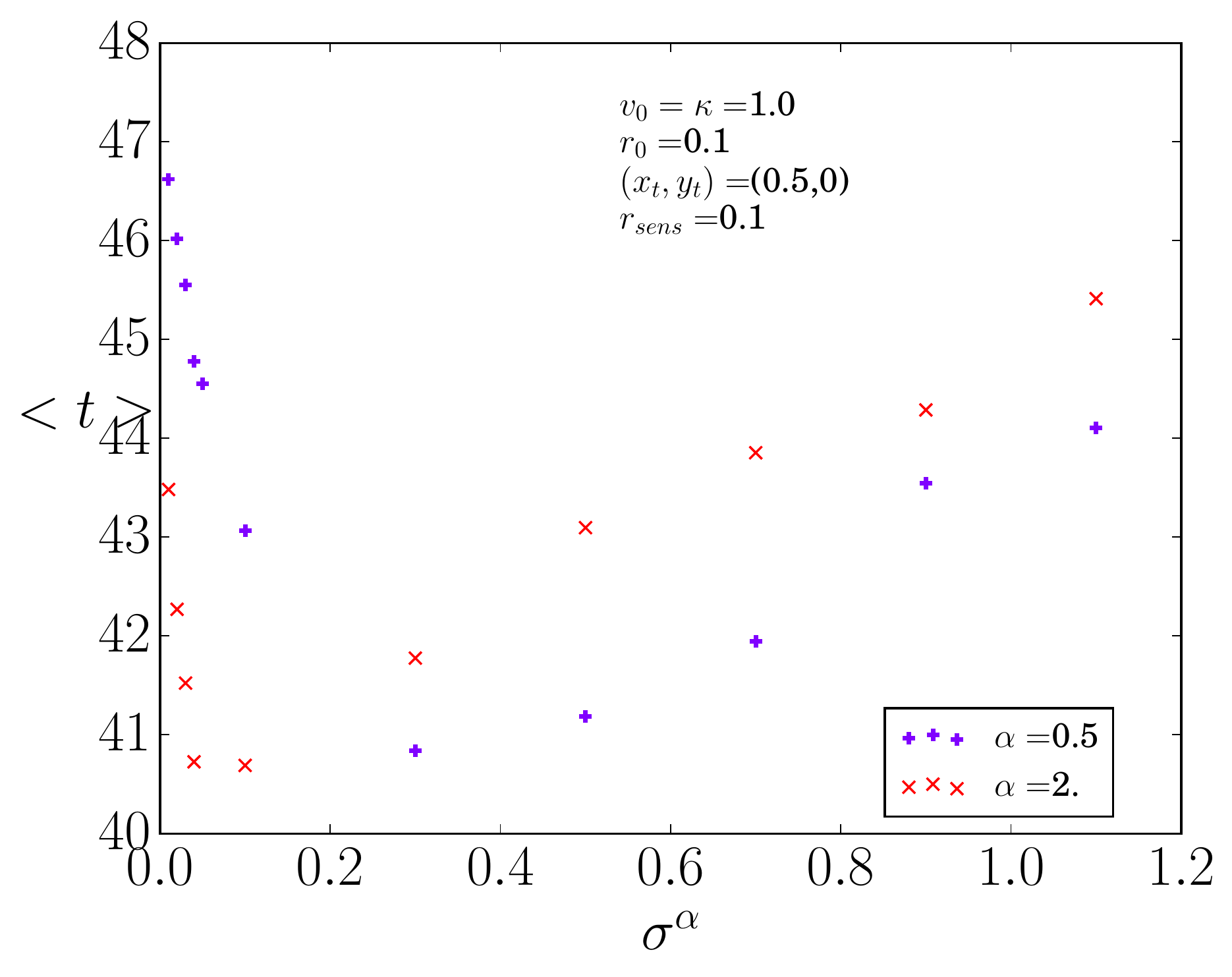}
    \includegraphics[width=0.49\linewidth]{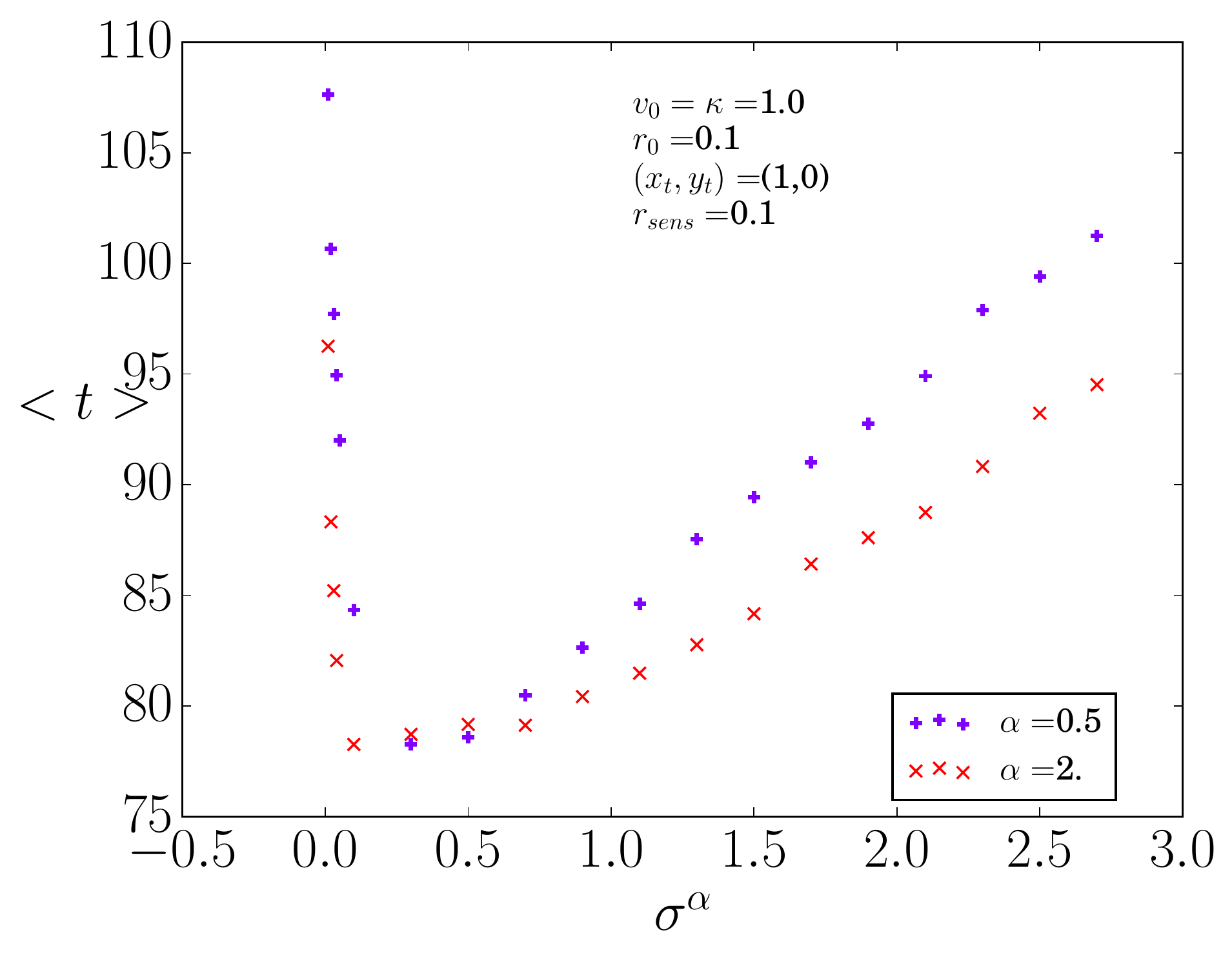}
    \caption{Mean hitting time $<t>$ during search of a new spot at
      two given positions with random reset at home.
      Left:$(x_t=0.5,y_t=0$);Right:$(x_t=1,y_t=0)$. The random reset of
      the searcher at a distance $r_{sens}=0.1$ from the home reduces
      the mean hitting times.}
    \label{fig:reset1}
\end{figure}

In Figures (\ref{fig:reset}) and (\ref{fig:reset1}) results of a new
situation are presented. There it was assumed that the searcher stays
between two sequential excursions a negligible period at the home. The
latter will not be counted but it shall be necessary that the
searcher forgets the incoming direction of the last search. Thus, it
starts after arrival the next search along an arbitrary direction
which was randomly selected. We believe that such restart from home is
closer for possible applications of food search.

Fig. (\ref{fig:reset}) shows again typical trajectories with a random reset
at the home. Fig. (\ref{fig:reset1}) presents the mean hitting times
of a possible target at distances $d=0.5$ and $d=1$. At the distance $d=1$ the marginal radial pdf is extremal 
for all noise types and values of the noise strength.  We
find a general reduction of the mean search time due to the additional
random effect.

\section{Searchers With Limited Knowledge Of The Position Angle}
\label{sec:extension_gamma}
The position angle $\beta$ influences the future heading direction as
given by equation \eqref{eq:dottheta}. In this section we investigate
the consequences of a limited knowledge of the position angle. A
limited knowledge means that the angle is not exactly known. To model this failure  we introduce a constant offset  $\gamma\in (-\pi/2,\pi/2)$ in the interaction rule of the heading and position vectors. That way we investigate in a simple way how the behavior of the searcher changes, if the understanding of the position angle is wrong. We will find, that even the slightest offset drastically changes the behavior of the particle, leading to the emergence of a limit cycle. The noise type as well as the noise strength however influence the pdf for being at the home and therefore for returning to it.
  
The time evolution of the heading direction $\theta$ becomes:
\begin{eqnarray}
&&\dot{x}= v_0 \cos(\theta)\nonumber\\
&&\dot{y}= v_0 \sin(\theta)\nonumber\\
&&\dot{\theta} =  \kappa\sin(\theta-\beta+\gamma) + \frac{\sigma}{v_0} \xi(t)\,.
\label{eq:dotthetagamma}
\end{eqnarray}
Like in the original model, we consider the coupling strength
$\kappa>0$ towards the home as constant.  For $\gamma=0$ this model
converges with the original model.  

Following Sect. \ref{sec:model}, we introduce dimensionless variables and change to polar coordinates. 
The time evolution of the position angle $\beta(t)$ remains unchanged as formulated in Eq.(\ref{eq:dim_less}). For the distance $r(t)$ and the
angle difference $z(t)=\theta(t)-\beta(t)$, we get
\begin{eqnarray}
&&\dot{r} = \cos(z) \nonumber \\
&&\dot{z} = \left(\cos(\gamma)-\frac{1}{r}\right)\sin(z)+ \sin(\gamma)\cos(z)+ \frac{\sigma}{v_0} \xi(t)
\label{eq:dotzgamma}
\end{eqnarray}
with $\xi(t)$ being $\alpha$ stable white noise. We consider again the wrapped angle $z\in(-\pi,\pi]$.

\subsection{The Deterministic Case}
Equation \eqref{eq:dotzgamma} shows that an uncertainty of the
position angle $\beta$ causes a change in the coupling strength
towards the home expressed through the $\cos(\gamma)$ term and also
causes an additional drift. Notably, for $|\gamma|=\pi/2$ the coupling
towards the home vanishes and for $|\gamma|>\pi/2$ the coupling
becomes repulsive. We restrict our discussion to
$|\gamma|<\pi/2$.

Unlike before, the deterministic reduced $(r,z)$ dynamics \eqref{eq:dim_less} $(\sigma=0)$ is now dissipative. In addition to the repelling zero-distance if $r(t) \rightarrow 0$, it has now one stable
and one unstable fixed points at
\begin{equation}
r_*=\frac{1}{\cos(\gamma)}\,
\label{eq:r_fix}
\end{equation}
and 
\begin{equation}
z_\pm=\pm\frac{\pi}{2}\,.
\end{equation}
 
Surprisingly, the steady $z_\pm$-values do not depend on $\gamma$. The
heading vector is always perpendicular to the position vector. In
contrast, the stationary distance from the home of the fixed points
increases with growing absolute value of $\gamma$ and reaches infinity
if $|\gamma|\rightarrow \pi/2$. The key reason for the transition from
the previous centers ($\gamma=0$) to a dissipative dynamics with
stable and unstable fixed points is the emergence of the term with
$\cos(z)$ on the r.h.s. of equation \eqref{eq:dotzgamma}. It is
proportional to the radial velocity $\dot{r}$ and vanishes for
$\gamma=0$. This term causes that small deviations in the vicinity of
$(r_*,z_\pm)$ either grow, or vanish.  It means, that eigenvalues of
the fixpoints for $\gamma \neq 0$ obtain a real part that is either
negative, or positive.

\begin{figure}
\begin{center}
    \includegraphics[width=0.49\linewidth]{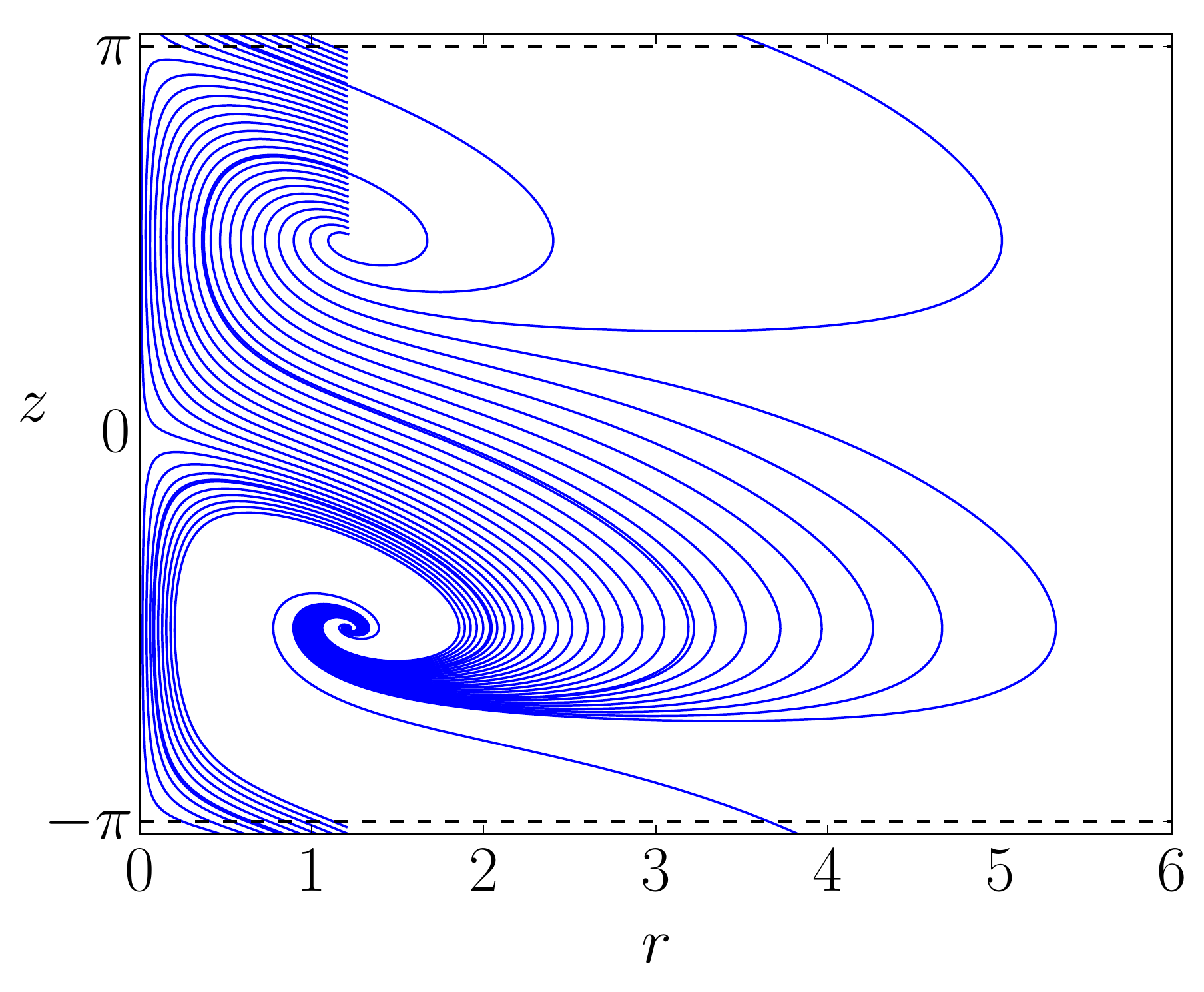}
    \includegraphics[width=0.49\linewidth]{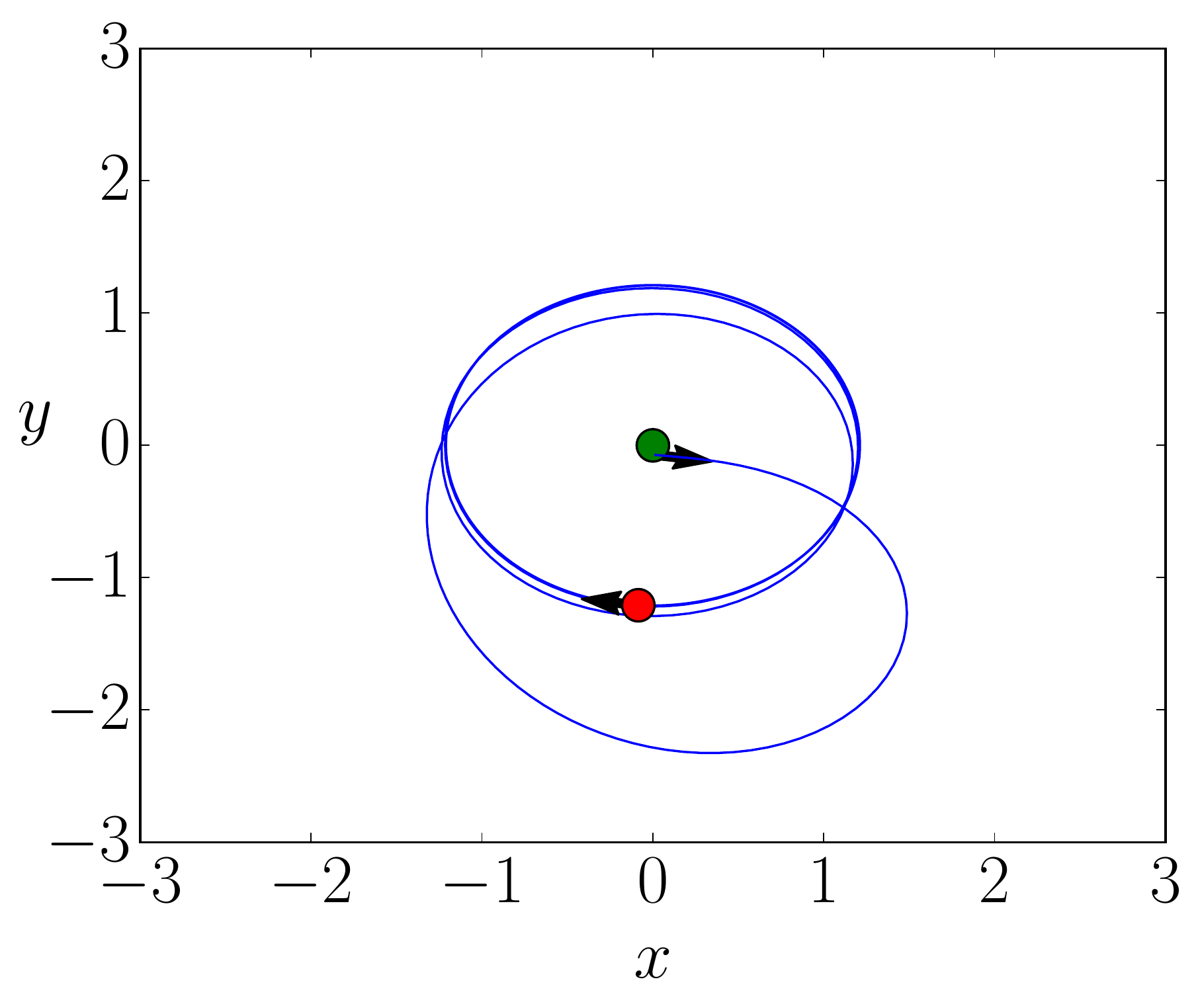}
\end{center}
    \caption{Left: Deterministic flow diagram of the dissipative wrapped dynamics in $r,z$ space with $\gamma=0.6$. Fix points are located at $r_*$ from Eq.(\ref{eq:r_fix}) and $z=\pm \pi/2$. Initial conditions $r_0=r_*$ and $z_0 \in [\pi/2,\pi]$. 
    The focus in the lower half plane is stable whereas the upper one is unstable. Right: Deterministic motion in the $x,y$ plane. Approaching the stable fixed point, 
    the searcher performs clockwise limit cycle oscillations in the $x,y$-space with heading vector perpendicularly to the position vector. }
    \label{fig:r_z_plane_gamma}
\end{figure}

In order to discuss the stability of the fixed points, we expand Eq. (\ref{eq:dotzgamma})  around the fixpoints $\{ r_*, z_\pm\}$  with small  $\delta_r=r-r_*$ and small $\delta_{z_\pm}=z-z_\pm$ up to first order of 
$\delta_r$ and $\delta_{z_\pm}$.  
The radial velocity becomes
\begin{equation}
\dot{\delta}_r =\mp\delta_{z_\pm} 
\label{eq:delta_r}
\end{equation}
and for the angle follows 
\begin{equation}
\dot{\delta}_{z_\pm} = \pm\cos^2(\gamma)\delta_r\mp\sin(\gamma)\delta_{z_\pm}\,\,.
\label{eq:delta_z}
\end{equation}
Taking another time derivative of the second equation leads to
\begin{equation}
 0=\ddot{\delta}_{z_\pm}   \pm \sin(\gamma)\dot{\delta}_{z_\pm} +\cos^2(\gamma)\delta_{z_\pm}\,\,
\label{eq:ddelta_z}
\end{equation}
the damped harmonic oscillator. The eigenvalues become, respectively,
$\lambda_\pm=\mp\sin(\gamma)/2\pm\sqrt{5\sin^2(\gamma)/4-1}$, meaning that for $\gamma\in(0,\pi/2)$ the fixed point $(r_*,z_+)$ is stable
and $(r_*,z_-)$ is unstable and vice versa for $\gamma\in(-\pi/2,0)$. It shall be also mentioned that values $\gamma \rightarrow \pi/2$ make the eigenvalues real and the foci becomes nodes.

The second term on the r.h.s. of (\ref{eq:ddelta_z}) is responsible for  damping. This term is a consequence of introducing $\gamma$.
For $\gamma=0$ the eigenvalues have no longer a real part, the second term on the r.h.s. vanishes, so $(r_*,z_\pm)$ become centers.

Figure \ref{fig:r_z_plane_gamma} shows trajectories for different initial conditions in the $(r,z)$ plane for $\gamma=0.6$. 
All trajectories converge to a single point $(r_*,z_-)$, except
the one started at the unstable fixed point $(r_*,z_+)$. 

Correspondingly sample trajectories in the $(x,y)$-plane converge to a circular oscillation with amplitude $r_*$ around the home  
in a clockwise fashion. The angular velocity follows from the $\beta$ dynamics $\dot{\beta}_*= -1/r_*=-\cos(\gamma)$. 
With positive $\gamma$ the $z_+$ becomes the stable fixpoint and the motion becomes anti clockwise around the home.  

\subsection{Small Noise Strength: Steady State - Gaussian White Noise}
\label{sec:dyn_gauss}
\begin{figure}
\begin{center}
    \includegraphics[width=0.49\linewidth]{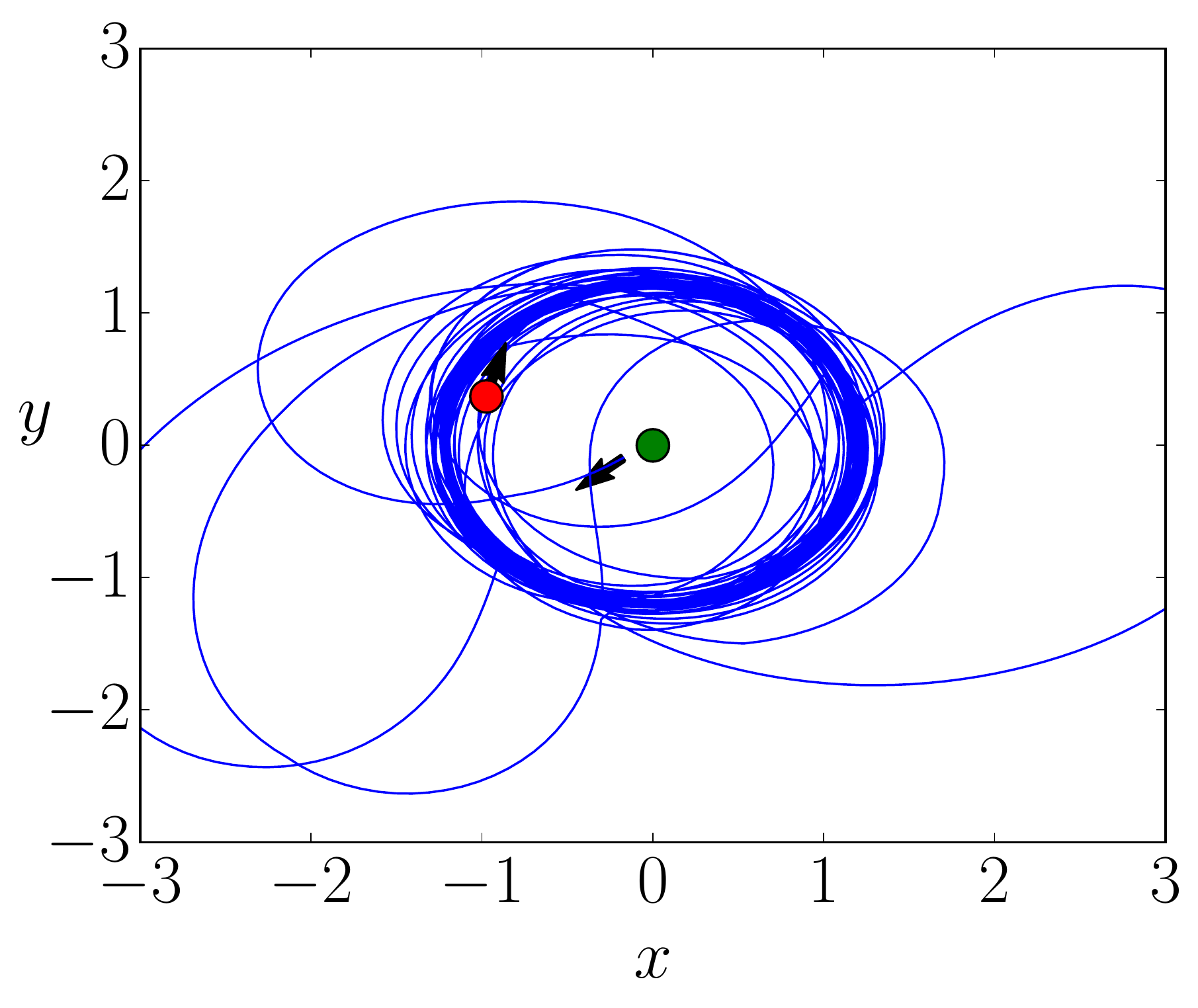}    \includegraphics[width=0.49\linewidth]{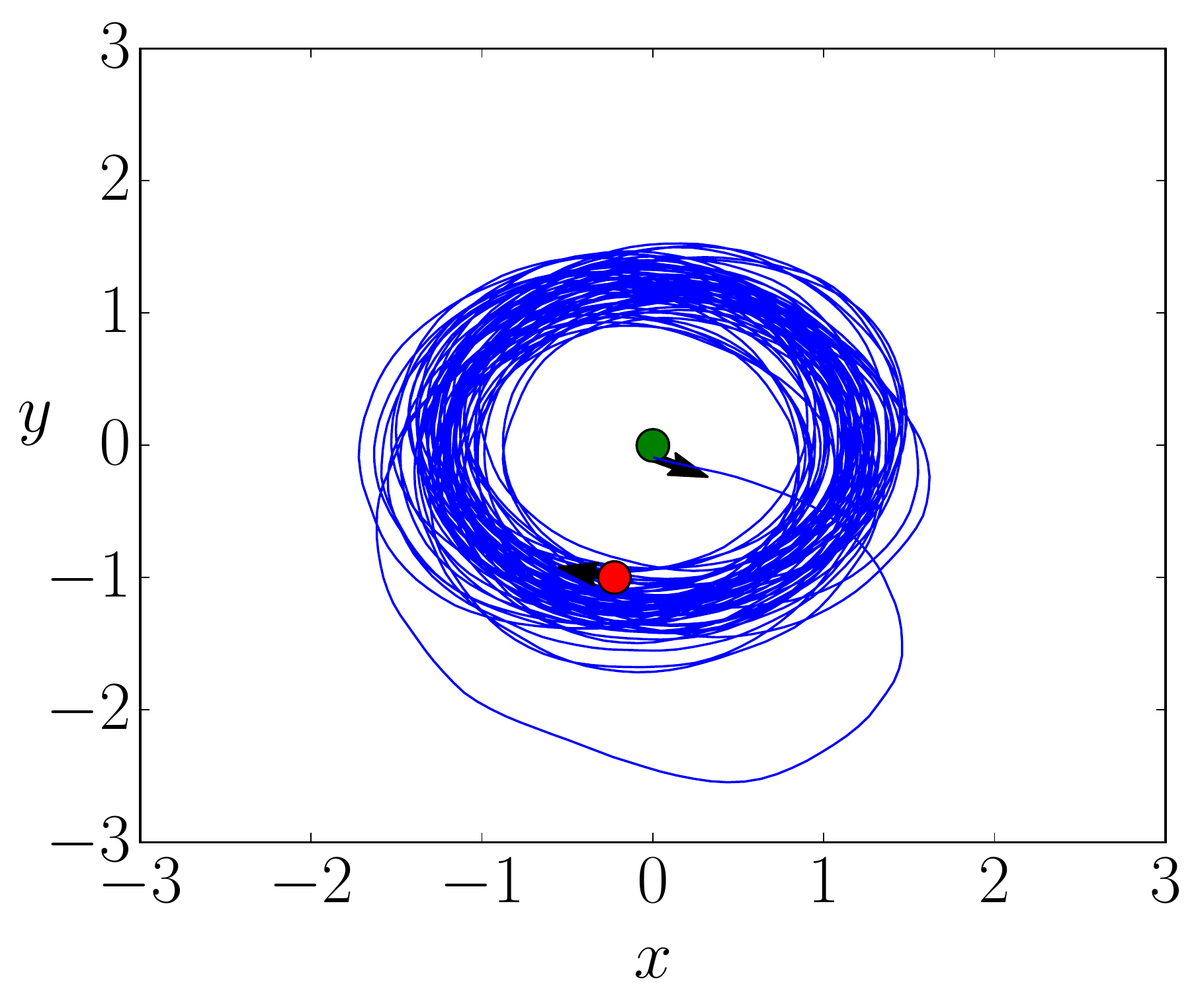}
\end{center}
    \caption{Stochastic trajectories of the searcher with shift $\gamma=-0.6$ in the interaction law starting at the home. Elapsed time during search: $t=500$. Left: $\alpha=1$, Right: $\alpha=2$. 
    Whereas the fluctuations around the limit cycle at the left are more narrow, the jumps in case of $\alpha=1$ excite larger deviations. Other parameters: $v_0=1$, $\sigma^\alpha=0.01$}
    \label{fig:stoch_gamma}
\end{figure}

Having briefly discussed the deterministic motion, we will now consider the noise in the heading dynamics. Typical stochastic trajectories are presented in Fig. (\ref{fig:stoch_gamma}) for small noise intensity $\sigma^\alpha=0.01$ and two values of $\alpha$. The left graph shows the searcher driven by white Cauchy noise with $\alpha=1$. In the situation presented at the right we took Gaussian white noise with $\alpha=2$. One immediately observes, that the motion is more narrow around the stable circular orbit in case of L{\'e}vy noise than in the Gaussian case. Otherwise, in the Gaussian case the motion is not interrupted by stronger deviations as the appear in the first case due to the discontinuous character of the L{\'e}vy noise. 

Next in subsection (\ref{sec:dyn_gauss1}), we will consider the application of Gaussian white noise in the heading-dynamics.  The other remaining parts (\ref{sec:dyn_alpha}) and (\ref{sec:alpha_large_noise})  will deal with situations that small and large $\alpha$-stable noise is applied. The distinction is originated by different approximations made for the aim to obtain expressions for the stationary spatial densities of the searchers.

\subsubsection{Small Noise Strength: Steady State - Gaussian White Noise}
\label{sec:dyn_gauss1}
As the investigation of the deterministic case has shown, the searchers move independently of initial conditions eventually 
on a circle of radius $r_*$ around the home in Cartesian coordinates. With noise added the distribution in the $x,y$ plane possesses circular symmetry. Hence, the marginal distribution of the distances is sufficient in order to characterize the asymptotic behavior of the searcher. Unfortunately, we have been unable to derive a general solution and will look for approximations, further on. Here,  we consider the linear equations \eqref{eq:delta_r} and \eqref{eq:delta_z}. We have to note, that the variables therein $\delta_r$ and $\delta_z$ have lost the
wrapped character. Both run between  $-\infty$ and $\infty$ and have lost the meaning of a positive distance and a $2\pi$ periodic angle. It implies that we have to restrict  the noise strength $\sigma$ to small values in the linear approximation. With small noise  one  avoids deviations where the character of a distance or periodicity becomes important.

Consider $\gamma$ to be positive. The stable fixed point is located in the upper $r,z$- halfplane. We formulate the stochastic dynamics $(\delta_r,\delta_z)$ around $(r_*, \pi/2)$ omitting the $\pm$ in the subscript of $\delta_z$. The corresponding FPE becomes
\begin{equation}
\frac{\partial}{\partial t}P(\delta_r,\delta_z,t)=\delta_z \frac{\partial}{\partial \delta_r}P-
\frac{\partial}{\partial \delta_z}\left(\cos^2(\gamma)\delta_rP-\sin(\gamma)\delta_zP\right)+\left(\frac{\sigma}{v_0}\right)^2\frac{\partial^2}{\partial (\delta_z)^2}P\,.
\label{eq:FPEdelta_r}
\end{equation}
We will look for its stationary solutions $P(\delta r,\delta z, t\rightarrow \infty) \rightarrow P_0(\delta r, \delta z)$ for $\gamma\in(0,\pi/2)$ and equate the l.h.s. to zero. This problem of a noise driven harmonic oscillator was already solved by Ornstein and Uhlenbeck \cite{Uhlenbeck}, the steady state pdf is given by
\begin{equation}
P_0(\delta_r,\delta_z)=\frac{v_0^2 \cos(\gamma)\,\sin(\gamma)}{2\pi\sigma^2} 
\exp\left(-\frac{v_0^2\sin(\gamma)\delta^2_z}{2\sigma^2}\right)\exp\left(-\frac{v_0^2\cos^2(\gamma)\sin(\gamma)\delta^2_r}{2\sigma^2}\right)\,.
\end{equation}
As explained above, the variables $\delta_z$ and $\delta_r$ are here considered to be not wrapped.

Reintroducing the position $r$ and the angle results in 
\begin{equation}
P_0(r,z)=\frac{v_0^2\cos(\gamma)\,\sin(\gamma)}{2\pi\sigma^2} 
\exp\left(-\frac{v_0^2\sin(\gamma)\left(z-\frac{\pi}{2}\right)^2}{2\sigma^2}
-\frac{v_0^2\cos^2(\gamma)\sin(\gamma)\left(r-\frac{1}{\cos(\gamma)}\right)^2}{2\sigma^2}\right)\,.
\end{equation}
Integration over the angle leads to the marginal distance 
dependent pdf $P_0(r)$:
\begin{equation}
P_0(r)=\frac{v_0 \cos(\gamma)\sqrt{\sin(\gamma)}}{\sqrt{2\pi}\sigma} 
\exp\left(-\frac{v_0^2\cos^2(\gamma)\sin(\gamma)\left(r-\frac{1}{\cos(\gamma)}\right)^2}{2\sigma^2}\right)\,.
\label{eq:marginal_gamma}
\end{equation}

\begin{figure}
\begin{center}
  \includegraphics[width=0.6\linewidth]{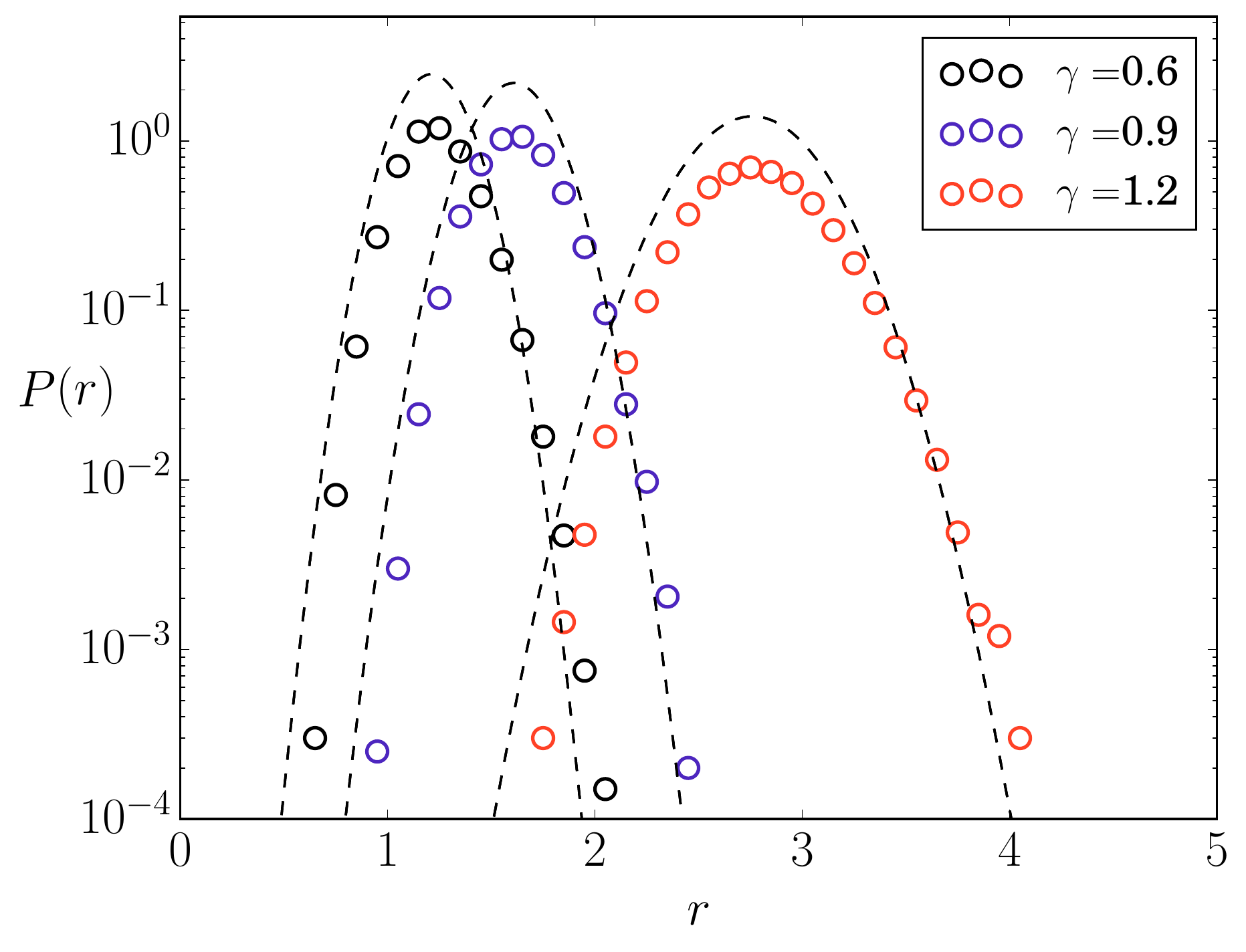}
\end{center}
    \caption{Steady state spatial pdf according to simulations of equations \eqref{eq:dotthetagamma} with \eqref{eq:beta} as symbols and theory given by \eqref{eq:marginal_gamma} as dashed lines. Parameter $\gamma$ given in the figure. Other parameters: $\sigma^2=0.01$, $v_0=1$, $\kappa=1$, $\alpha=2$.}
    \label{fig:FPE_gamma}
\end{figure}
This marginal pdf is compared with simulation results in figure
\ref{fig:FPE_gamma}. The colored symbols are obtained from simulations
according to equations \eqref{eq:dotthetagamma},
with the definition of the position angle $\beta$, given by
\eqref{eq:beta}. The black dashed lines correspond to equation
\eqref{eq:marginal_gamma}. The approximations
\eqref{eq:marginal_gamma} fit the simulation results well for the
small noise intensity $\sigma^2=0.01$. 

We remind that we selected $\gamma \in (0,\pi/2)$. A corresponding distribution can be derived with negative shifts if the stable fixpoint is located in the lower halfplane. In the Cartesian coordinate system both possess the same steady state pdf as circular symmetric crater-like distribution with radius $r_*$. The crater is accompanied by a homogeneous rotation around the home in an
either clockwise $\gamma<0$, or counterclockwise $\gamma>0$ fashion.

\subsubsection{Small Noise Strength: Steady State - $\alpha-$Stable White Noise}
\label{sec:dyn_alpha}
In this subsection we will discuss situations with $\alpha-$stable
white noise sources being not Gaussian. This distinction seems
appropriate for two reasons. First, the expansion for small $\delta_z$
is questionable for Non-Gaussian $\alpha$-stable white noise, as those
processes are not continuous, they jump.  So, an approximation for
small $\delta_z$ does not consider jumps, it has to be done with
caution. We do it anyway and will justify this by an agreement with
simulation results, as presented later on. As we infere from the left
graphs of Fig.(\ref{fig:stoch_gamma}), jumps are rare and lead to
large excursions, so the approximation can still be usefull at the
center of the pdf.  Secondly, as was shown in \cite{harm_osz_sok} the
position and the velocity are coupled in a non trivial way for
harmonic oscillators driven by L\'evy noise, in general. Only for
Gaussian white noise the resulting pdf can be written as product of
the angular dynamics and the position dynamics \cite{harm_osz_sok}. We
will find an approximation only for the marginal pdf of the position
and then discuss the result in comparison with simulations.

Despite these initial warnings, we start the same as in the Gaussian
white noise case with the dynamics obtained by expansion for small
$\delta_z$ and $\delta_r$ given by \eqref{eq:delta_r} and
\eqref{eq:delta_z}.  We write both equations as second derivative of
the position $ \delta_r$ with noise present
\begin{equation}
 0=\ddot{\delta}_r+\sin(\gamma)\dot{\delta}_r
 +\cos^2(\gamma)\delta_r+\sigma\xi(t)\,,
\label{eq:ddelta_r}
\end{equation}
for $\gamma>0$.  We consider the overdamped regime. We look at time
scales $t\gg(\sin(\gamma))^{-1}$ and eliminate the acceleration, leaving with
\begin{equation}
 \dot{\delta}_r=-\frac{1}{\sin(\gamma)}\left(\cos^2(\gamma) \delta_r+\frac{\sigma}{v_0} \xi(t)\right)\,.
\label{eq:delta_r_adi}
\end{equation}
The noise is symmetric, so we can keep the plus sign in front of the
noise term.  Such equation \eqref{eq:delta_r_adi} was solved and
discussed in \cite{west}. We take their result for the asymptotic $t
\rightarrow \infty$ and return to the coordinate $r$ and express the
steady state spatial pdf for the distance through the Fourier
transform
\begin{equation}
P_0(r)=\int_{-\infty}^\infty\,{\rm{d}}k\,\exp\left(-ik\left(r-\frac{1}{\cos(\gamma)}\right) \right)\exp\left(-\left(\frac{\sigma}{v_0}\right)^\alpha \frac{\sin(\gamma)}{\alpha\sin^\alpha(\gamma)\cos^2(\gamma)}|k|^\alpha \right)\,.
\label{eq:marginal_gamma_all}
\end{equation}
For the special case of Cauchy distributed white noise with $\alpha=1$ the result
becomes
\begin{equation}
  P_0(r)=\frac{\sigma}{\pi v_0}\frac{1}{\left(
    r\cos(\gamma)-1\right)^2+\left( \sigma/(v_0\cos(\gamma))\right)^2}
\label{eq:marginal_gamma_cauchy}
\end{equation}
for the marginal steady state spatial density.

\begin{figure}
\begin{center}
  \includegraphics[width=0.48\linewidth]{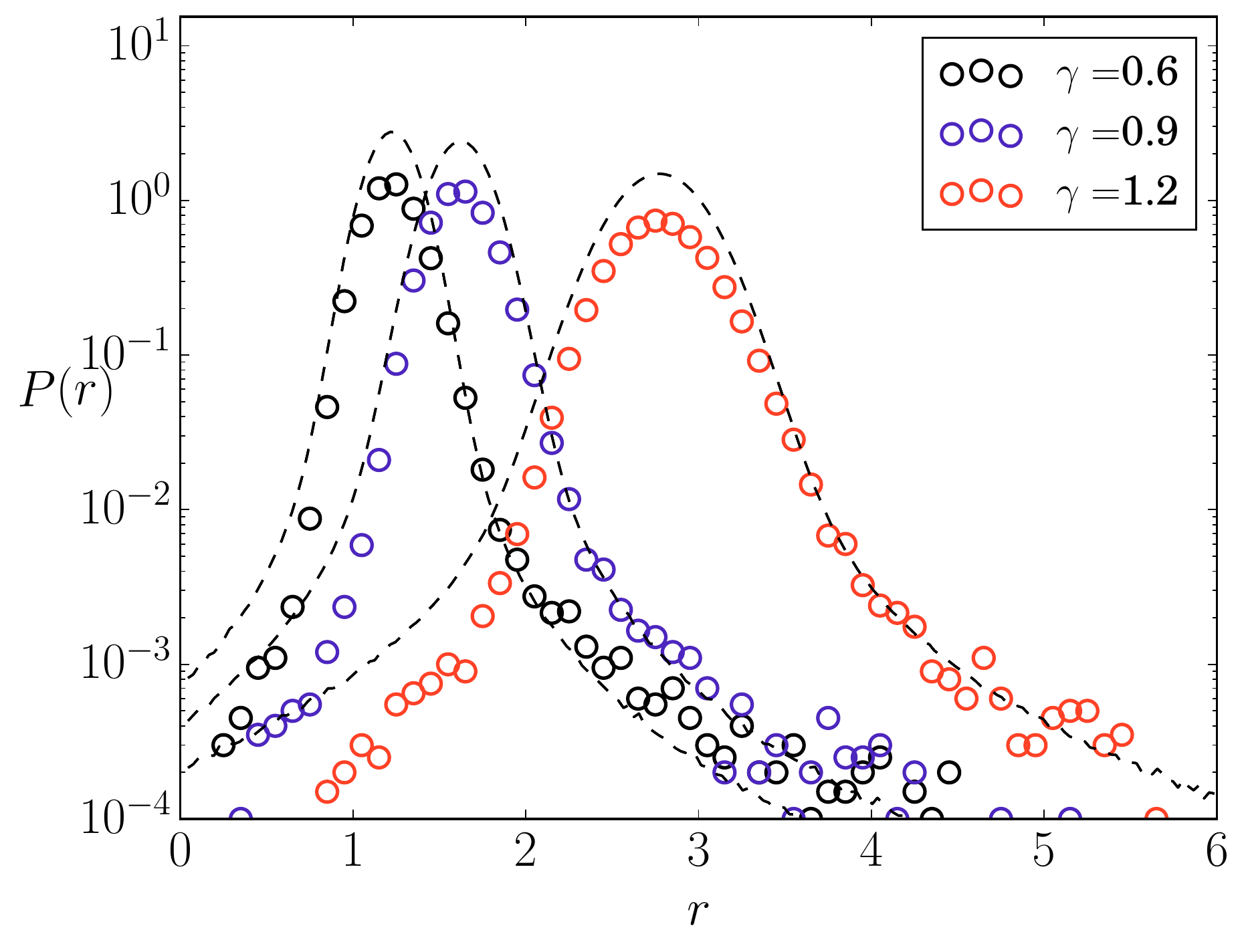}
  \includegraphics[width=0.48\linewidth]{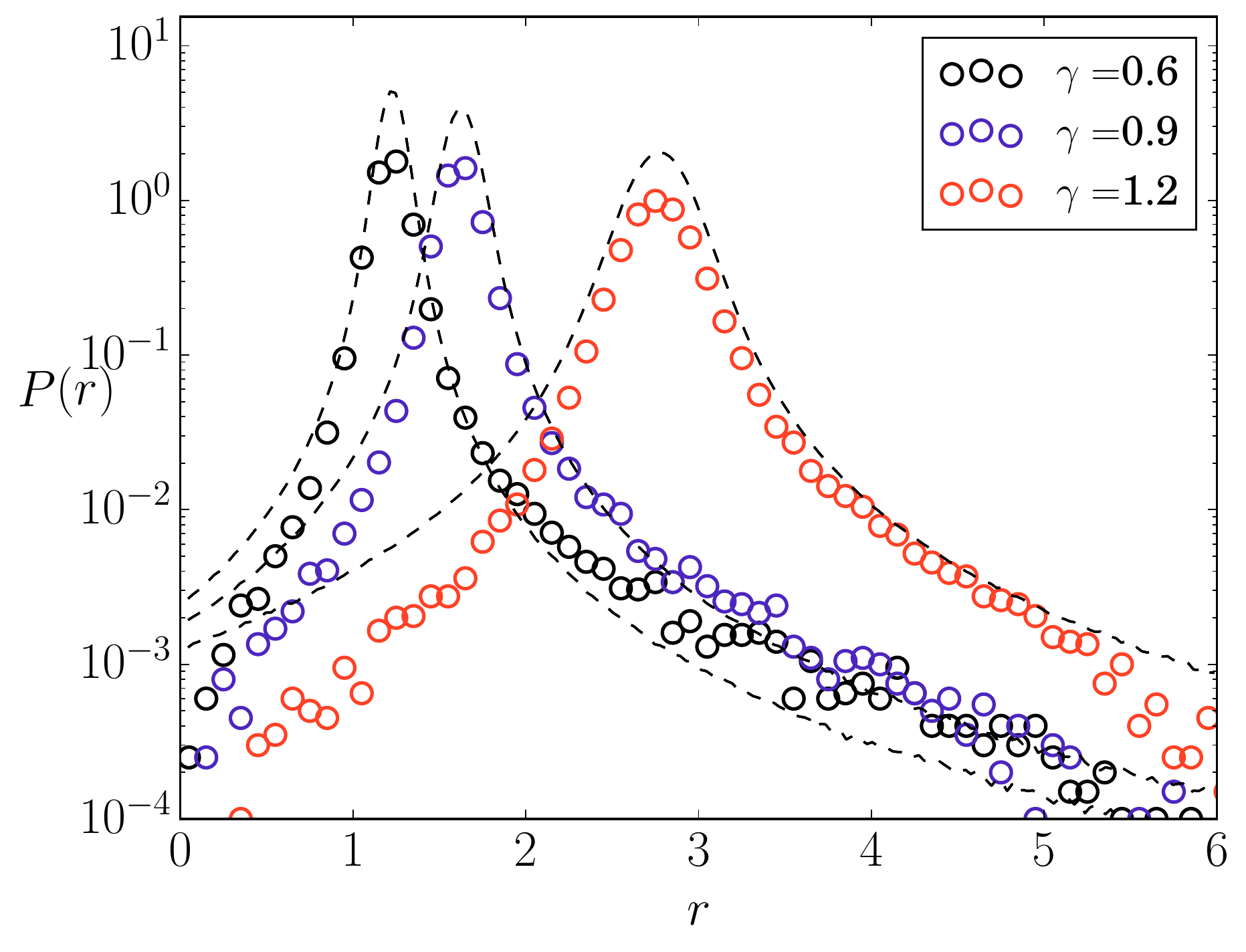}
  \includegraphics[width=0.48\linewidth]{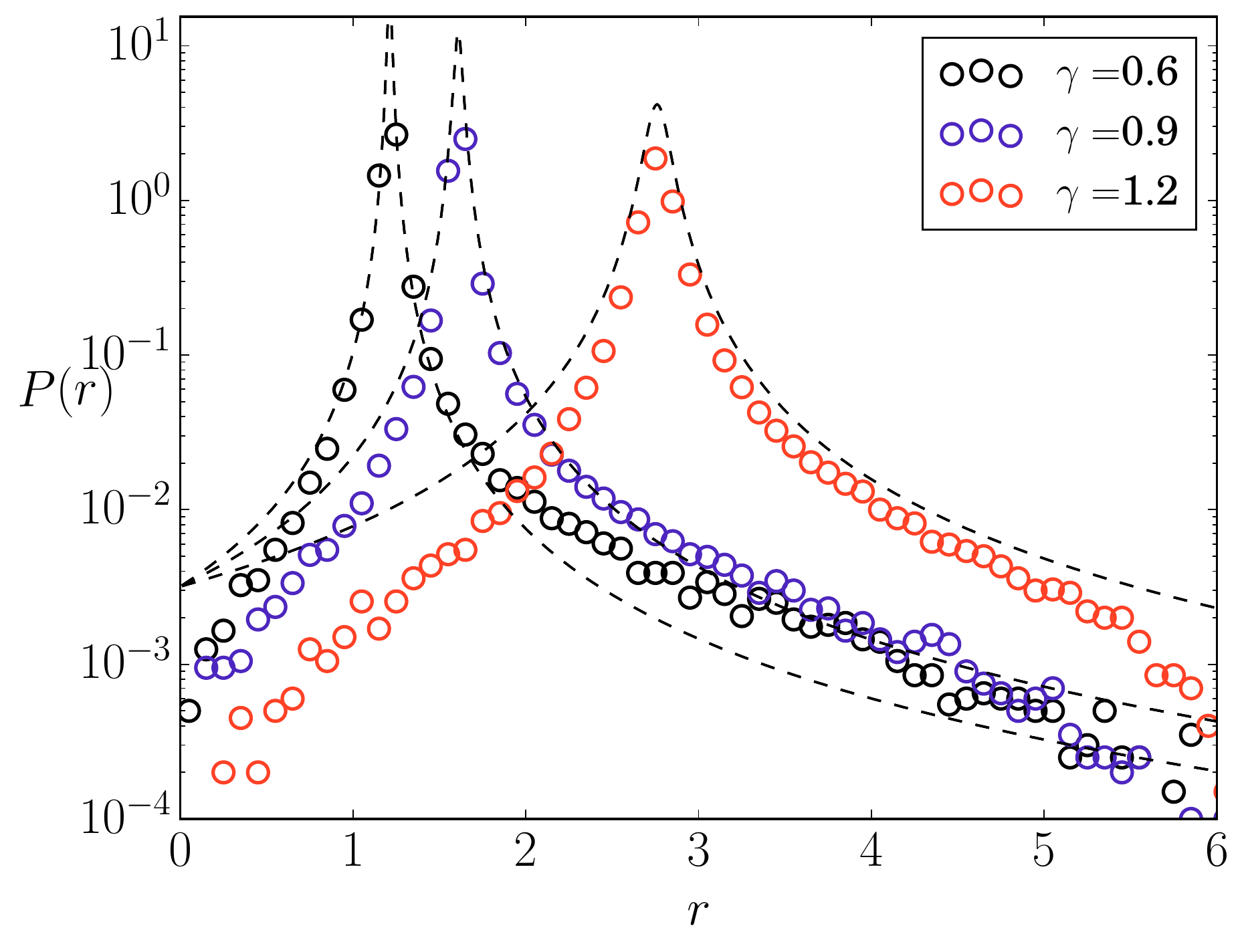}
  \includegraphics[width=0.48\linewidth]{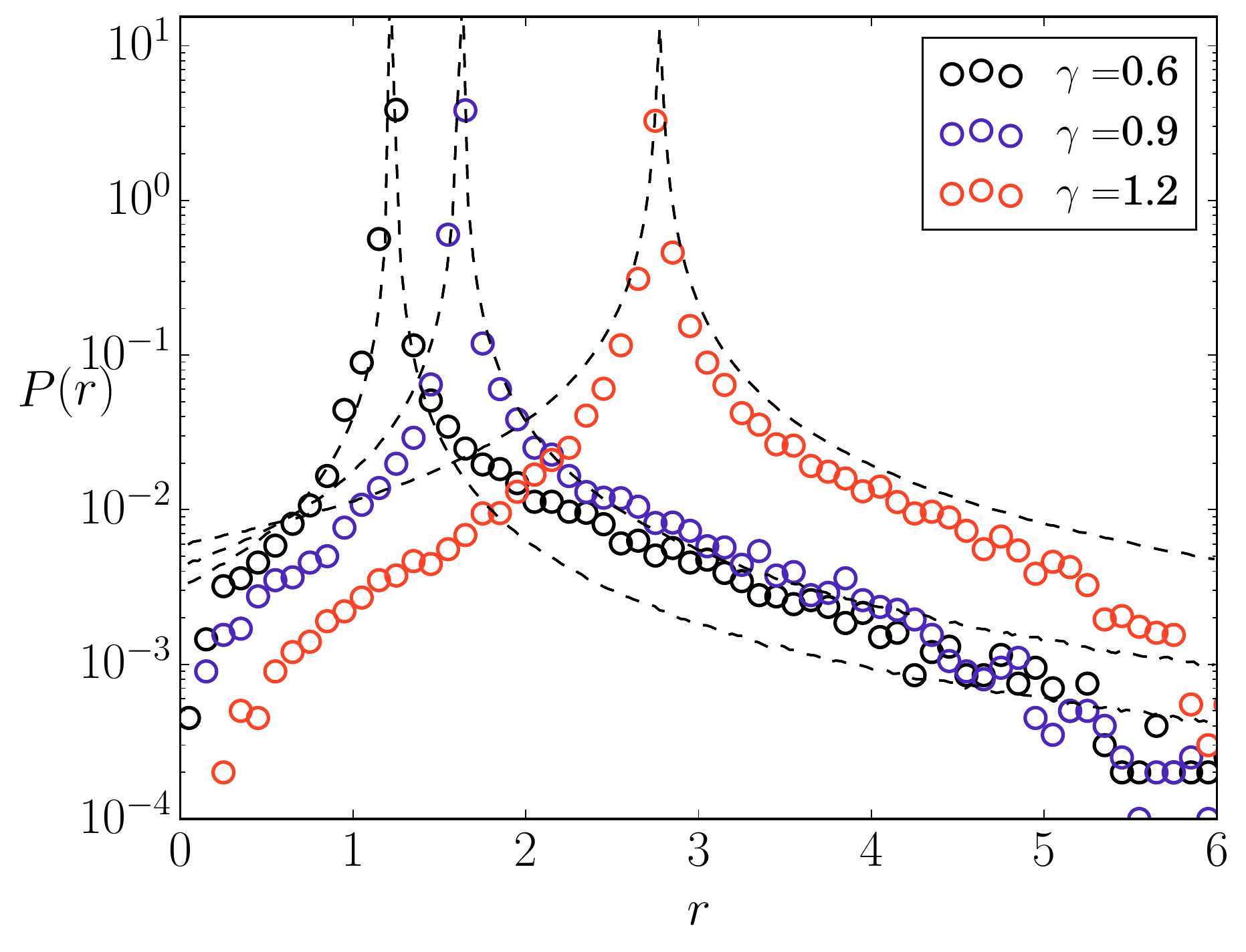}
\end{center}
    \caption{Marginal steady state spatial pdf according to
      simulations of equations \eqref{eq:dotthetagamma} as symbols  with $\beta$ defined in \eqref{eq:beta} and
      theory given by \eqref{eq:marginal_gamma} as dashed lines. Top
      left: $\alpha=1.9$, Top right: $\alpha=1.5$, Bottom left:
      $\alpha=1.0$, Bottom right: $\alpha=0.5$. Parameter $\gamma$
      given in the figure. Other parameters: $\sigma^\alpha=0.01$,
      $v_0=1$.}
    \label{fig:FPE_gamma_cauchy}
\end{figure}
We compare in Fig. (\ref{fig:FPE_gamma_cauchy}) the approximations
\eqref{eq:marginal_gamma_all}, \eqref{eq:marginal_gamma_cauchy} with
simulation results. The colored symbols are obtained from simulations according to
equations \eqref{eq:dotthetagamma}, with the
definition of the position angle $\beta$ given by
\eqref{eq:beta}. The black dashed lines correspond to equations
\eqref{eq:marginal_gamma_all}, \eqref{eq:marginal_gamma_cauchy}.  The
parameter $\gamma$ is given in the figures. The noise strength was
chosen to be small $\sigma=0.01$.  We took for the simulations $v_0=1$.  We display in the top row the cases $\alpha=1.9$
(left) and $\alpha=1.5$ (right) and in the bottom row $\alpha=1.0$
(left) and $\alpha=0.5$ (right).  While decreasing $\alpha$ the peak
at $r_*$ becomes sharper and the tails become longer. The
approximations fit generally rather well around the peak, but are
better for larger $\alpha$ and smaller $\gamma$ values. For larger
$\alpha$ values less jumps occur, so the approximation is expected to
work better. As we expanded for small $\delta_r$ around $r_*$, the
approximation does not reflect the tails correctly. A smaller $\gamma$
value corresponds to a stronger force around $r_*$, so if the angular
variable experiences a jump and therefore the particle moves away from
$r_*$, it returns faster for small $\gamma$, or slower for larger
$\gamma$.  This way it can be understood that the approximations work
better for larger $\alpha$ and smaller $\gamma$.

Focussing on the simulation results of Figs. (\ref{fig:FPE_gamma}) and (\ref{fig:FPE_gamma_cauchy}), we underline here, that the results of the pdf show close to the home higher values with decreasing $\alpha$, meaning that for Gaussian white noise the particle practically never is close to home (for instance $r=0.2$), while it has a larger probability density with $\alpha=0.5$. While we found that the mean first hitting time of a food source
does not significantly depend on the noise type, in case the position angle is exactly known, we find here that for the return to the home part the noise type 
might matter, if an uncertainty of the position angle $\beta$ exists. Therefore a specific turning behavior of an observed animal might be related to the accuracy of its understanding of the surroundings and not to optimization of the search itself.

We remind that the distance $r$ was initially derived as polar
representation of a two dimensional Cartesian system. As in the case
of Gaussian noise, we ignored that distances can not be negative. In
the simulations the density $P_0(r)$  approaches zero for vanishing
$r$. It differs in our approximation and probability
is also found at vanishing $r$ and below. Nevertheless, although some
steps to derive the approximation \eqref{eq:marginal_gamma_all} have
to be taken with caution, the obtained approximations fit the
simulations around the maximal values rather well and differences to
the Gaussian cases become obvious.

Likewise in the Gaussian case, in the Cartesian coordinate system the
steady state pdf for the distance is accompanied by a symmetric
rotation around the home in an either clockwise $\gamma<0$, or
counterclockwise $\gamma>0$ fashion.

\subsubsection{Large Noise Strength}
\label{sec:alpha_large_noise}
In the subsection, we derive the steady state spatial density for
large noise strength $\sigma$. We will eliminate higher orders of the
relaxation time $\tau=(v_0/\sigma)^\alpha$, as this time scale vanishes
for $\sigma\rightarrow \infty$.

We start from the equation of motion for the dimensionless dynamics \eqref{eq:dotzgamma} containing $\gamma$ and wherein  the
position angle $\beta(t)$ is due to \eqref{eq:beta}. Outgoing from this stochastic Langevin equation we formulate the 
corresponding FPE
\begin{equation}
\frac{\partial}{\partial t}P=-\cos(z)\frac{\partial}{\partial
  r}P-\frac{\partial}{\partial z}
\left[\left(\cos(\gamma)-\frac{1}{r}\right)\sin(z)+\sin(\gamma)\cos(z)\right]P+\left(\frac{\sigma}{v_0}\right)^\alpha
\frac{\partial}{\partial |z|^\alpha}P\,,
\label{eq:FPE_gamma_highfull}
\end{equation}
for the pdf $P=P(r,z,t|r_0,z_0,t_0)$. We express the latter one through the
angular $2 \pi$ Fourier-transform with complex amplitudes
\begin{equation}
P_n(r,t)=\frac{1}{2\pi}\int_{-\pi}^{\pi} {\rm d}z\, \exp(-inz)P(r,z,t)\,.
\nonumber
\end{equation}
Multiplying the FPE \eqref{eq:FPE_gamma_highfull} from the left with
$\exp(\text{i}nz)$ and integration over the angular variable $z$ leads to a set of linear equations for the Fourier amplitudes. We obtain the hierarchy:
\begin{eqnarray}
\frac{\partial}{\partial t}P_n&&=-\frac{1}{2}\frac{\partial}{\partial
  r}\left(P_{n+1}+P_{n-1}\right)+\frac{n}{2}
\left(\cos(\gamma)-\frac{1}{r}\right)\left(P_{n+1}-P_{n-1}\right)+\nonumber\\ &&+\frac{\text{i}n}{2}\sin(\gamma)\left(P_{n+1}+P_{n-1}\right)
-\left(\frac{\sigma}{v_0}\right)^\alpha |n|^\alpha P_n\,.
\end{eqnarray}

Being interested in the steady state, we set the l.h.s to zero, i.e. $\partial P_n/\partial t=0$, leaving:
\begin{eqnarray}
\frac{|n|^\alpha}{\tau} P_n&&=-\frac{1}{2}\frac{\partial}{\partial r}\left(P_{n+1}+P_{n-1}\right)+\frac{n}{2}
\left(\cos(\gamma)-\frac{1}{r}\right)\left(P_{n+1}-P_{n-1}\right)+\nonumber\\
&&+\frac{\text{i}n}{2}\sin(\gamma)\left(P_{n+1}+P_{n-1}\right)\,,
\label{eq:FPE_gamma_highfull_k}
\end{eqnarray}
with $\tau$ being the relaxation time from \eqref{eq:tau}.

Further on, we find an approximative solution for large noise $\sigma$, respectively, for small $\tau$. The amplitudes with $n=0,\pm1$ obey the two relations:
\begin{eqnarray}
0=-\frac{1}{2}\frac{\partial}{\partial r}\left(P_{1}+P_{-1}\right)\,,
\label{eq:FPE_gamma_p0}
\end{eqnarray}
\begin{eqnarray}
P_{\pm1}&&=-\frac{\tau}{2}\frac{\partial}{\partial r}\left(P_{\pm2}+P_{0}\right)
+\frac{\tau}{2}
\left(\cos(\gamma)-\frac{1}{r}\right)\left(P_{\pm2}-P_{0}\right)+\nonumber\\
&&\pm\frac{\text{i}\tau}{2}\sin(\gamma)\left(P_{\pm2}+P_{0}\right)\,,
\label{eq:FPE_gamma_highfull_k1}
\end{eqnarray}
Insertion of $P_{\pm1}$ into the first equation \eqref{eq:FPE_gamma_p0} yields
\begin{eqnarray}
0&&=-\frac{\tau}{4}\frac{\partial}{\partial r}\left\{\frac{\partial}{\partial r}2P_{0}+2\left(\cos(\gamma)-\frac{1}{r}\right)P_0+ \right. \nonumber \\
&&\left.+\left(\frac{\partial}{\partial r}-\cos(\gamma)+\frac{1}{r}\right)\left(P_{2}+P_{-2}\right)
-i\sin(\gamma)\left(P_{2}-P_{-2}\right)\right\}\,.
\label{eq:FPE_gamma_highfull_k0}
\end{eqnarray}
The functions $P_{\pm2}$ are of order $\tau P_0$. These functions become negligible as $\tau\rightarrow 0$. Hence, we find for large noise the steady state spatial density as:  
\begin{equation}
P_0(r)=\cos^2(\gamma)r\exp(-\cos(\gamma)r)\,.
\label{eq:FPE_gamma_high}
\end{equation}
And thus, for large noise strength the steady state pdf becomes independent of the noise type and asymptotically independent of the noise strength $\sigma$.


\begin{figure}
\begin{center}
  \includegraphics[width=0.6\linewidth]{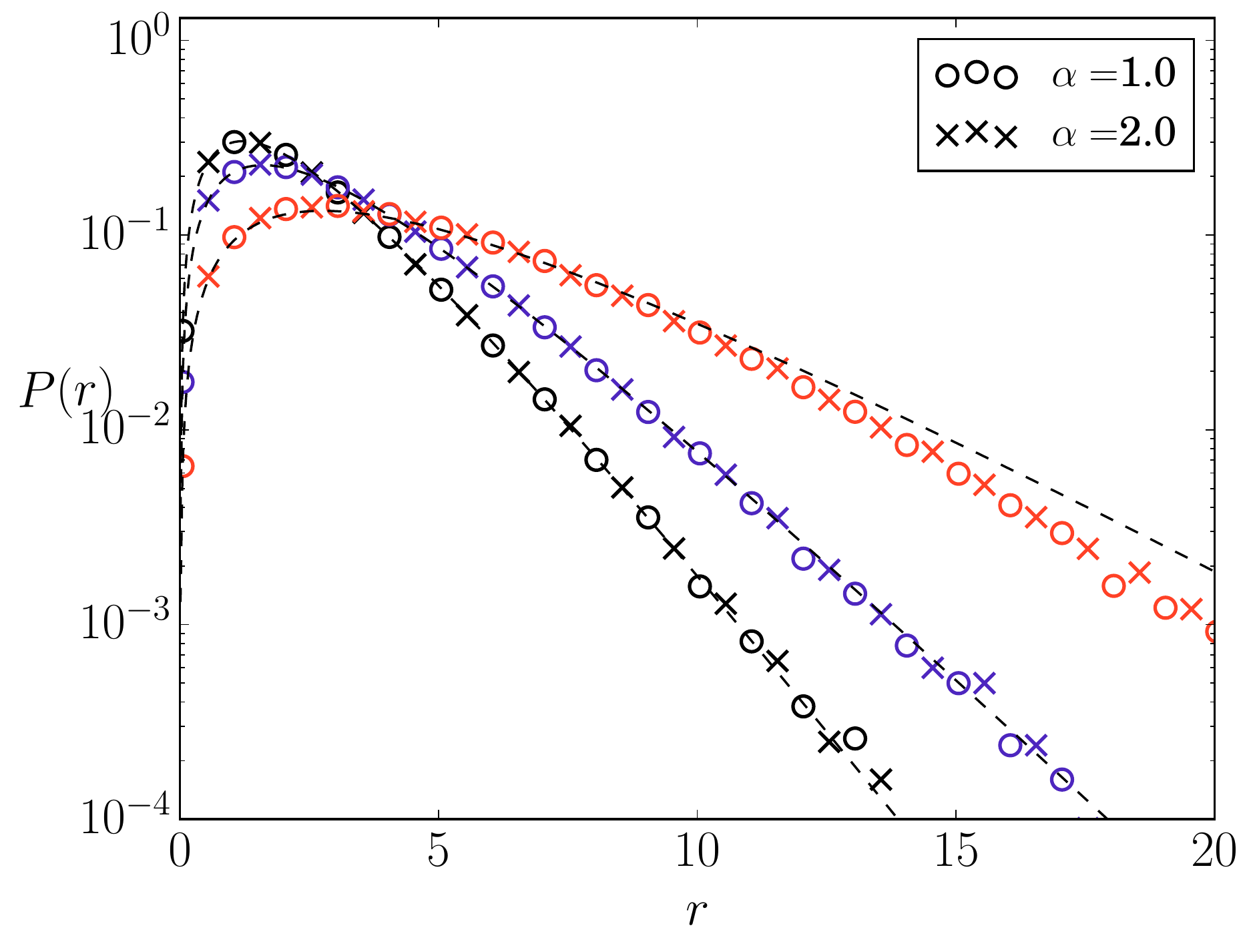}
\end{center}
    \caption{Marginal steady state spatial densities according to simulations with  as symbols at high noise strength $\sigma^\alpha=4.0$. Symbol 'o' corresponds to Cauchy distributed white noise $\alpha=1$
    and symbol 'x' to Gaussian white noise. The dashed lines are 
    according to equation \eqref{eq:FPE_gamma_high}. Parameter $\gamma=0.6$ (black), $\gamma=0.9$ (blue) , $\gamma=0.9$ (red).
    Other  parameters: $v_0=1$, $\kappa=1$.}
    \label{fig:FPE_gamma_high}
\end{figure}

Figure \ref{fig:FPE_gamma_high} shows simulation and theoretical
results. Here, simulation results for Gaussian white noise (symbol x)
and Cauchy distributed white noise (symbol o) are plotted for three
different values of $\gamma$ and $\sigma^\alpha=4$.  The marginal
density remains dependent on $\gamma$ but does no longer depend on
$\alpha$. The dashed lines in Fig. (\ref{fig:FPE_gamma_high}) correspond to the approximative solution (\ref{eq:FPE_gamma_high}). As can be seen,  they fit the simulation results rather well.

This result is somewhat suprising. We introduced the shift $\gamma$,
as an uncertainty of the position angle. We found that for small noise, 
the noise type can significantly increase the value of the pdf
close to the home, the point where the particle wishes to return to.
Now, we find that increasing the noise strength, and therefore
changing the heading directions rather frequently drastically
increases the pdf close to the home. As important result it implies the chances of
returning to the home are significantly increased, as the running in
circles motion is interrupted.

\section{Conclusions}
\label{sec:concl}
We resumed the study on a recently proposed stochastic model for a
local searcher which is bound to a certain position called home
\cite{Noetel_2018}. The search around and the return to the home is
described by an uniform rule. It bases on an escape and pursuit
interaction of the heading and the position vector.  If both vectors
point into the same direction the heading vector repels from the
position vector in order to explore more space. Oppositely, if the
heading vector points homewards, the two vectors align in order to find the
home.

The model was composed with these properties to explain recent experimental findings  with food searching fruit flies which perform stochastic oscillatory search around a given food position \cite{Kim_Dickinson_2017}. In a recent publication 
\cite{Noetel_2018} we showed the qualitative agreement with the experiment. 
However, the proposed model might be useful for much more applications of animal motion, for example, 
for explaining the search pattern of desert ants \cite{Wehner_1981,Wehner_et_al_1996,Vickerstaff}, of general search \cite{Klages} as well 
as technical applications like self-navigation of robots and social situations as mushrooming and orientation of visitors in unknown places 
and clients of supermarkets to find the exit.    

Another main ingredient of the model was the constant speed $v_0$ of
the searcher. We added noise only in the turning angle behavior as
observed in the experiments with the fruit fly. Different white noise
types as Gaussian and other stable L\'evy noise was used as source of
the randomness in the decision making of selecting a new heading
direction. Interestingly, all different noise types with varying
intensity did not have influence on the spatial density of searchers
around the home.

Oppositely, the characteristic measure in the model which is affected by the noise is the mean time of finding a new  localized food spot at a certain distance from the home. We reported on an optimal noise strength for finding this new spot
in a minimal mean time $<t>$. It appeared to be the consequence of two counteracting effects driven by the noise. On the one side increasing noise populates stronger different orbits with probability. On the other side, a strong noise shrinks the influence of the deterministic motion which becomes replaced by low diffusion.

This optimal average time is distance
dependent. The searcher finds on average the second spot always faster
with noise in the angular dynamics.  This is the result of the
relaxation towards a probabilistic population of all possible
trajectories which determines the greater success of the stochastic
searcher. For lower noise this process is governed by the noisy
periodic motion and after the relaxation time the stationary pdf is
established. However, for larger noise the relaxation is proceeded by
diffusive search.

In the second part, we discussed consequences of an erroneous
observation of the position angle $\beta$ by the active particle. We
did so by introducing an offset $\gamma$ in the interaction of heading and position vectors. We found that the resulting
motion becomes circular around the home in Cartesian coordinates with the
radius given by equation \eqref{eq:r_fix} if $|\gamma| \in (0,\pi/2)$.
We obtained, that the spatial pdf depends on the noise type and its strength in case of  small intensities. We  approximated the spatial densities by linearizing the deterministic drift and discussed differences between white Gaussian and Cauchy $\alpha$-stable noise. In case of large noise again a noise independent distribution for the stochastic distances of the circular motion around the home was found.

We found, that while the mean first hitting time of a food source does
not significantly depend on the noise type, that either noise type
(for small noise strength), or noise strength (independent of the
noise type) can increase the stationary pdf close to the home, if an
uncertainty of the position angle exists. This finding might suggest
that a specific in experiments observed turning behavior can be rooted
in an uncertainty of the position angle and not in optimization of the
time for finding a food source.

It would be more realistic to introduce time dependent process
$\gamma(t)$ mimicking the forgetting of the actual direction of the
home due to continued small errors. Such work is in progress. Also
interacting searchers and their cooperative behavior are of central
interest in future research.

\section{Acknowledgments}
This work was supported by the Deutsche Forschungsgemeinschaft via
grant IRTG 1740 and by the Sao Paulo Research Foundation (FAPESP) via
grants 2015/50122-0 and 2017/04552-9. LSG thanks Dr. Alexander Neiman
and Ohio University in Athens OH for hospitality and support.  The
authors thank Fabian Baumann and Bartlomiej Dybiec for fruitful discussions.

\appendix
\section{Numerical Integration}
\label{app:numeric}
The symmetric stable random variable $X$,with scale parameter $\sigma=1$, 
can be generated from uniform distributed random numbers $U_1, \,U_2\in(0,1)$ \cite{Weron_book, Nolan}, by
\begin{eqnarray}
V&&=\pi(U_1-1/2)\nonumber\\
W&&=-\log(U_2)\nonumber\\
X&& = \left\{
\begin{matrix} 
\frac{\sin{(\alpha V)}}{\cos{(V)^{1/\alpha}}} \left[  \frac{\cos{((\alpha-1)V)}}{W} \right]^{(1-\alpha)/\alpha}   & , \alpha \neq 1\\
\tan{(V)} & , \alpha = 1
\end{matrix}
\right.
\label{eq:rand_stable}
\end{eqnarray}
Considering Eq.\eqref{eq:dottheta}, we perform the numeric integration by a deterministic Euler step with 
additional noise 
\begin{equation}
\theta(t+\Delta t)=\kappa \sin(\theta-\beta)\Delta t + \frac{\sigma}{v_0}X_1\cdot\Delta t^\frac{1}{\alpha}+\theta(t)\,\,
\label{eq:num_lange_stable_def}
 \end{equation}
with $X_1$ being a random variable drawn from a stable distribution. Due to the large tales in the pdf for the noise, we take a time step 
of $\Delta t =10^{-4}$.

\section{Fractional Derivative of Sine Functions}
\label{app:sin}
We show here, that the symmetric $\alpha$th Riesz-Weyl derivative for a sine function is the sine function with a pre-factor, see also \cite{Noetel_pre}. 
Considering the function $f(z)=\sin(b\,z)$, with $b$ being a real number and $z$ to be unwrapped $z\in(-\infty,\infty)$,
the fractional derivative becomes
\begin{equation}
\frac{\partial^\alpha}{\partial |z|^\alpha}f(z)=-\frac{1}{4\,{\rm{i}}\,\pi}\int_{-\infty}^\infty\,{\rm{d}k}|k|^\alpha\,\exp\left(-ikz\right)\,\left(\delta(k+b)-\delta(k-b) \right)\,
\end{equation}
where we included the Fourier transform of the sine function. 
Evaluating the $k$ integration leads to:
\begin{equation}
\frac{\partial^\alpha}{\partial |z|^\alpha}f(z)=-|b|^\alpha\,\sin(b\,z)\,.
\end{equation}

%

\end{document}